\documentclass[reprint,aps,prd,twocolumn,superscriptaddress,showpacs,preprintnumbers,amsmath,amssymb,floatfix,nofootinbib]{revtex4-1}%

\pdfoutput=1

\usepackage{graphicx}
\usepackage{epstopdf}
\usepackage[mathscr]{euscript}
\usepackage{mathtools} 
\usepackage{textcomp} 
\usepackage{dcolumn}
\usepackage[caption=false]{subfig}
\usepackage{bigstrut}
\usepackage[english]{babel}
\usepackage{bm}
\usepackage[mathlines]{lineno}
\usepackage{xspace} 
\usepackage{multirow}
\usepackage[colorlinks,citecolor=blue]{hyperref}
\usepackage[dvipsnames]{xcolor}
\usepackage{enumitem}
\newcommand{\FIG}{Fig.~}

\newcommand{\SEC}{Sec.~}
\newcommand{\SECS}{Secs.~}
\newcommand{\TAB}{Table~}

\newcommand{\EQ}{Eq.~}
\newcommand{\EQS}{Eqs.~}
\newcommand{\REF}{Ref.~}

\newcommand{\Ereeeq}{E_{\text{r,ee}}\xspace} 
 
\newcommand{\Ernreq}{E_{\text{r,nr}}\xspace} 
\newcommand{\Ernr}{$E_{\text{r,nr}}$\xspace} 
\newcommand{\Ereq}{E_{\text{r}}\xspace} 
\newcommand{\Er}{$E_{\text{r}}$\xspace} 
\newcommand{\Eteq}{E_{\text{t}}\xspace} 

\newcommand{\ELeq}{E_{\text{NTL}}\xspace}

\def\evee{eV$_\text{ee}$\xspace}

\def\kevt{keV$_\text{t}$\xspace}

\def\kevee{keV$_\text{ee}$\xspace}
\def\kevnr{keV$_\text{nr}$\xspace}

\def\keveeeq{\text{keV}_\text{ee}\xspace}

\def\epgeq{\epsilon_{\gamma}}

\def\cm2{cm$^\text{2}$\xspace}

\def\gev{GeV/$c^2$\xspace}

\def\cf{$^{252}$Cf\xspace}
 
\newcommand{\runOne}{Run~1\xspace}
\newcommand{\runTwo}{Run~2\xspace}
\newcommand{\runThree}{Run~3\xspace}
\newcommand{\perOne}{Period~1\xspace}
\newcommand{\perTwo}{Period~2\xspace}
\newcommand{\cdmsII}{CDMS\,II\xspace}
\newcommand{\scdms}{SuperCDMS\xspace}
\newcommand{\geant}{G\textsc{eant}4\xspace}

\newcolumntype{d}[1]{D{,}{.}{#1}}
\newcolumntype{p}[1]{D{,}{\,\pm\,}{#1}}
\newcolumntype{h}[1]{D{,}{\text{--}}{#1}}

\begin{document}
\title{Search for low-mass dark matter with CDMSlite using a profile likelihood fit}

\affiliation{Division of Physics, Mathematics, \& Astronomy, California Institute of Technology, Pasadena, CA 91125, USA}
\affiliation{Department of Physics, Durham University, Durham DH1 3LE, UK}
\affiliation{Fermi National Accelerator Laboratory, Batavia, IL 60510, USA}
\affiliation{Lawrence Berkeley National Laboratory, Berkeley, CA 94720, USA}
\affiliation{School of Physical Sciences, National Institute of Science Education and Research, HBNI, Jatni - 752050, India} 
\affiliation{Department of Physics \& Astronomy, Northwestern University, Evanston, IL 60208-3112, USA}
\affiliation{Pacific Northwest National Laboratory, Richland, WA 99352, USA}
\affiliation{Department of Physics, Queen's University, Kingston, ON K7L 3N6, Canada}
\affiliation{Department of Physics, Santa Clara University, Santa Clara, CA 95053, USA}
\affiliation{SLAC National Accelerator Laboratory/Kavli Institute for Particle Astrophysics and Cosmology, Menlo Park, CA 94025, USA}
\affiliation{SNOLAB, Creighton Mine \#9, 1039 Regional Road 24, Sudbury, ON P3Y 1N2, Canada}
\affiliation{Department of Physics, South Dakota School of Mines and Technology, Rapid City, SD 57701, USA}
\affiliation{Department of Physics, Southern Methodist University, Dallas, TX 75275, USA}
\affiliation{Department of Physics, Stanford University, Stanford, CA 94305, USA}
\affiliation{Department of Electrical and Computer Engineering, Texas A\&M University, College Station, TX 77843, USA}
\affiliation{Department of Physics and Astronomy, and the Mitchell Institute for Fundamental Physics and Astronomy, Texas A\&M University, College Station, TX 77843, USA}
\affiliation{Instituto de F\'{\i}sica Te\'orica UAM/CSIC, Universidad Aut\'onoma de Madrid, 28049 Madrid, Spain}
\affiliation{D\'epartement de Physique, Universit\'e de Montr\'eal, Montr\'eal, Qu\'ebec H3C 3J7, Canada}
\affiliation{Department of Physics \& Astronomy, University of British Columbia, Vancouver, BC V6T 1Z1, Canada}
\affiliation{Department of Physics, University of California, Berkeley, CA 94720, USA}
\affiliation{Department of Physics, University of Colorado Denver, Denver, CO 80217, USA}
\affiliation{Department of Electrical Engineering, University of Colorado Denver, Denver, CO 80217, USA}
\affiliation{Department of Physics, University of Evansville, Evansville, IN 47722, USA}
\affiliation{Department of Physics, University of Florida, Gainesville, FL 32611, USA}
\affiliation{School of Physics \& Astronomy, University of Minnesota, Minneapolis, MN 55455, USA}
\affiliation{Department of Physics, University of South Dakota, Vermillion, SD 57069, USA}
\affiliation{Department of Physics, University of Toronto, Toronto, ON M5S 1A7, Canada}
\affiliation{TRIUMF, Vancouver, BC V6T 2A3, Canada}

\author{R.~Agnese} \affiliation{Department of Physics, University of Florida, Gainesville, FL 32611, USA} 

\author{T.~Aralis} \affiliation{Division of Physics, Mathematics, \& Astronomy, California Institute of Technology, Pasadena, CA 91125, USA} 

\author{T.~Aramaki} \affiliation{SLAC National Accelerator Laboratory/Kavli Institute for Particle Astrophysics and Cosmology, Menlo Park, CA 94025, USA} 

\author{I.J.~Arnquist} \affiliation{Pacific Northwest National Laboratory, Richland, WA 99352, USA} 

\author{E.~Azadbakht} \affiliation{Department of Physics and Astronomy, and the Mitchell Institute for Fundamental Physics and Astronomy, Texas A\&M University, College Station, TX 77843, USA} 

\author{W.~Baker} \affiliation{Department of Physics and Astronomy, and the Mitchell Institute for Fundamental Physics and Astronomy, Texas A\&M University, College Station, TX 77843, USA} 

\author{S.~Banik} \affiliation{School of Physical Sciences, National Institute of Science Education and Research, HBNI, Jatni - 752050, India} 

\author{D.~Barker} \affiliation{School of Physics \& Astronomy, University of Minnesota, Minneapolis, MN 55455, USA} 

\author{D.A.~Bauer} \affiliation{Fermi National Accelerator Laboratory, Batavia, IL 60510, USA} 

\author{T.~Binder} \affiliation{Department of Physics, University of South Dakota, Vermillion, SD 57069, USA} 

% ...

\smallskip

\date{\today}

\collaboration{SuperCDMS Collaboration}
\noaffiliation

\smallskip

\begin{abstract}

The Cryogenic Dark Matter Search low ionization threshold experiment (CDMSlite) searches for interactions between dark matter particles and germanium nuclei in cryogenic detectors. The experiment has achieved a low energy threshold with improved sensitivity to low-mass ($<$10\,GeV/$c^2$\xspace) dark matter particles. We present an analysis of the final CDMSlite data set, taken with a different detector than was used for the two previous CDMSlite data sets.  This analysis includes a data ``salting'' method to protect against bias, improved noise discrimination, background modeling, and the use of profile likelihood methods to search for a dark matter signal in the presence of backgrounds. We achieve an energy threshold of 70~eV and significantly improve the sensitivity for dark matter particles with masses between 2.5 and 10\,GeV/$c^2$\xspace compared to previous analyses.  We set an upper limit on the dark matter-nucleon scattering cross section in germanium of 5.4$\times$10$^{-42}$\,cm$^2$ at 5\,GeV/$c^2$\xspace, a factor of $\sim$2.5 improvement over the previous CDMSlite result.  

\end{abstract}


\maketitle


\section{Introduction}
\label{sec:intro}

Multiple astronomical and cosmological observations point to 
the existence of dark matter (DM), indicating that approximately 25\% of the universe consists of a non-luminous, non-baryonic form of matter of unknown composition~\cite{Tanabashi2018,Ade2016}. 

A class of hypothetical particles called Weakly Interacting Massive Particles (WIMPs) \cite{STEIGMAN1985} is consistent with the observational evidence and would be a cold (non-relativistic) relic from the early universe that may be directly detectable by terrestrial detectors~\cite{Jungman1996}.

Supersymmetric theories naturally predict the existence of WIMPs with masses at the electroweak scale, but with no evidence of such particles at the LHC~\cite{Aad2015,Khachatryan2015}, direct-detection DM experiments have begun to consider low-mass alternatives~\cite{Aguilar-Arevalo2016,Petricca2017,Agnese2018,Agnes2018}. Theories that predict DM particles with masses $\lesssim 10$\,GeV/$c^2$ include, but are not limited to, asymmetric DM, which relates the DM problem to the baryon asymmetry of the universe~\cite{Petraki2013,Zurek2014}, and hidden sector scenarios in which DM couples to Standard Model particles through new force mediators like the dark photon~\cite{Hooper2012,Foot2013}. 

In CDMSlite, cryogenic germanium detectors developed by the SuperCDMS Collaboration were operated at high voltage to amplify the signal from ionization by particle interactions via the Neganov-Trofimov-Luke (NTL) effect~\cite{Neganov1985,Luke1988}. This amplification provides sub-keV detection thresholds for nuclear recoils, enabling searches for low-mass DM particles ~\cite{Agnese2014a,Agnese2016,Agnese2018}.
This paper presents results from  the third and final run of CDMSlite, and represents the first blind analysis of data taken in this mode. We employ new rejection techniques to effectively remove instrumental backgrounds that limited previous analyses, while the remaining dominant background contributions are modeled within a profile likelihood fit.

Section \ref{sec:expDescription} describes the operation and calibration of CDMSlite detectors. 
Section \ref{sec:salting} presents a method of data blinding based upon the addition of artificial events to the data, while \SEC\ref{sec:dataAnalysis} describes how instrumental backgrounds are effectively removed. Section~\ref{sec:fiducialvolume} describes the definition of a fiducial volume (using the radial parameter discussed in \SEC\ref{sec:expDescription}) to eliminate the contribution of events with misreconstructed energies at high detector radii.  Sections~\ref{sec:sigEff} and \ref{sec:backgrounds} discuss models for the energy spectra of DM-signal and background events, which are used as inputs to a profile likelihood fit to search for a DM signal in \SEC\ref{sec:likelihood}.  We find no evidence for such a signal and present improved upper limits on the spin-independent DM-nucleon cross section in \SEC~\ref{sec:results}.
\section{Description of the Experiment}
\label{sec:expDescription}

The \scdms Soudan experiment was located at the Soudan Underground Laboratory in northern Minnesota. The experiment operated  15 germanium interleaved Z-sensitive Ionization and Phonon (iZIP) detectors, arranged in 5 stacks (``towers'') and read out with \cdmsII electronics~\cite{Akerib2004,Akerib2005,Ahmed2009,Ahmed2009}. 
The iZIPs---cylindrical Ge single crystals with a diameter of $\sim$76\,mm, a height of $\sim$25\,mm, and a resulting mass of $\sim$600\,g---were equipped with phonon sensors composed of tungsten transition edge sensors (TESs) and aluminum fins for phonon collection, patterned on their top and bottom faces. The operational temperature was $\sim$50\,mK. Interleaved with the phonon sensors were charge-collecting electrodes with a bias voltage applied between them (+2\,V on one face and -2\,V on the other) to separate and collect the electrons and holes liberated in particle interactions. Nuclear recoils (NRs) produce fewer electron-hole pairs for a given recoil energy than electron recoils (ERs), allowing for an event-by-event discrimination between these two types of interactions~\cite{Agnese2013}.

In 2012 we explored the operation of an iZIP detector in an alternative configuration ~\cite{Agnese2014a} in which a higher bias across the detector amplifies the ionization signal by producing NTL phonons. As charge carriers drift across the crystal due to the electric field, they quickly reach a terminal velocity and the additional work done on the carriers is transferred to the crystal lattice in the form of NTL phonons. The energy contribution from NTL phonons is
\begin{equation}
	\ELeq = e \, \Delta V N_{\text{e/h}},
    \label{eq:E_ntl}
\end{equation}
where $e$ is the absolute value of the electric charge, $\Delta V$ the voltage drop experienced by a charge pair, and $N_{\text{e/h}}$ is the number of electron-hole pairs produced. The total phonon energy generated by a recoiling particle is the sum of the initial recoil energy $\Ereq$ and the energy of NTL phonons:
\begin{equation}
	\Eteq = \Ereq + \ELeq.
    \label{eq:totEne_noYield}
\end{equation}
\noindent
In germanium, the average energy required to produce one electron-hole pair for an electron recoil is $\epgeq = 3.0$~eV~\cite{Antman1966}, giving $N_{\text{e/h}} = \Ereq/\epgeq$. Therefore, a 75\,V potential difference across the detector amplifies the ionization signal by a factor of 26 for an electron recoil.

The hardware trigger, based on the total phonon signal, was tuned on the CDMSlite detector to achieve as low of a threshold as possible while also maintaining a manageable trigger rate. The hardware trigger threshold, measured using the method described in \SEC IV.B of \REF\cite{Agnese2018}, varied approximately between 50 and 70~eV, which resulted in trigger rates between 0.2 and 15~Hz over the course of the run. When a trigger occurs, the data acquisition electronics record the phonon signals as waveforms, digitized at 625\,kHz and lasting $\sim$6.6\,ms, from all active detectors in the array. The signals from the charge-collecting electrodes were also read out as waveforms for each trigger; however, this information was only used to remove events with particularly bad noise in the charge waveforms.

\subsection{Energy Scale}
\label{sec:expDescription:energyScale}
``Electron equivalent'' energy units (\kevee) are the most convenient for analysis of data from the CDMSlite runs, because the observed backgrounds consist primarily of ER events. The electron equivalent energy is the electron recoil energy that would produce the same amount of phonon energy as is observed in the detector.

We calibrate the energy scale using a $^{252}$Cf neutron source.  Activation of $^{70}$Ge by neutron capture produces $^{71}$Ge, which decays by electron capture with a 11.43 day half-life~\cite{Hampel1985}. These decays produce peaks at the $K$-, $L$-, and $M$-shell binding energies of $^{71}$Ga of 10.37, 1.30, and 0.16~keV, respectively~\cite{Bearden1967}. The prominent $K$-shell peak is used to calibrate the energy scale to \kevee and correct for any time variation in the detector response. The corrections and calibration were found to be appropriate for the less prominent $L$- and $M$- shell peaks, indicating detector response linearity throughout the energy range of interest.

NRs produce fewer charge pairs and therefore a smaller ionization signal than ERs of the same recoil energy, and we parametrize the smaller ionization signal by the energy-dependent ionization yield $Y(\Ereq)$. The number of electron-hole pairs is then given by $N_{\text{e/h}} = Y(\Ereq)\Ereq/\epgeq$.  The total measured energy in terms of the event recoil energy and ionization yield is:
\begin{equation}
	\Eteq = \Ereq \left( 1 + Y(\Ereq)\frac{e\Delta V}{\epgeq} \right).
    \label{eq:totEne_yield}
\end{equation}
For ERs, $Y \equiv 1$ by definition. To convert from an electron equivalent energy to a nuclear-recoil equivalent energy (denoted \Ernr with units of \kevnr), we correct \EQ\ref{eq:totEne_yield} for the difference in yield between nuclear and electron recoils, while assuming that each electron-hole pair experiences the full applied bias $V_{\textrm{det}}$:
\begin{equation}
	\Ernreq = \Ereeeq \left( \frac{1 + e V_{\textrm{det}}/\epgeq}{1 + Y(\Ernreq) e V_{\textrm{det}}/\epgeq} \right).
    \label{eq:NRene}
\end{equation}

We use the Lindhard model~\cite{Lindhard1963a,*Lindhard1963,*Lindhard1968} for the yield as a function of nuclear-recoil energy: 
\begin{equation}
	Y(\Ernreq) = \frac{k \cdot g(\varepsilon)}{1 + kg(\varepsilon)},
    \label{eq:lindhard}
\end{equation}
where $g(\varepsilon) = 3\varepsilon^{0.15} + 0.7\varepsilon^{0.6} + \varepsilon$, $\varepsilon = 11.5\Ernreq(\text{keV})Z^{-7/3}$, and $Z$ is the atomic number of the detector material. Measurements of $Y$ in germanium are generally consistent with a small range of $k$ values approximately centered on the Lindhard model prediction of $k = 0.157$~\cite{Jones1971,*Jones1975,Sattler1966,Messous1995,Barbeau2007}. We account for the spread in experimental measurements as a systematic uncertainty on $k$, as discussed in \SEC\ref{ref:likelihood:systematics}.

\subsection{Operating Conditions}
\label{sec:expDescription:operations}

For CDMSlite \runThree, we operated a single detector in CDMSlite mode from February to May 2015 for a total livetime of 60.9~days. The top detector in the second tower was selected based on the two qualities that contributed most to lower analysis thresholds. First, this detector exhibited stable operation for a range of applied bias voltage up to nearly 75\,V. Second, because of its reduced susceptibility to vibrational noise, this detector's phonon energy resolution was among the best in the detector array.  While a different detector was selected for CDMSlite \runOne and \runTwo based on the same two metrics, the decision to switch detectors for \runThree was also intended to demonstrate reproducibility of the CDMSlite operating technique across multiple detectors.

We applied a 75~V bias to one side of the detector with the other side grounded, following the same biasing scheme used in the previous CDMSlite runs. We also adopted the \runTwo ``pre-biasing'' procedure in which the detector bias was temporarily increased (to 85 V for \runThree) prior to the start of each data series~\cite{Agnese2018}.

The voltage at the detector differed from the applied voltage $V_b$ because of a parasitic resistance that caused a significant voltage drop across a bias resistor ($R_b =196$~M$\Omega$) upstream of the detector. The parasitic resistance caused a current draw from the power supply, $I_{\textrm{HV}}$, which we continuously measured in order to monitor the detector voltage: 
\begin{equation}
	V_{\textrm{det}} = V_{b} - I_{\textrm{HV}}R_{b}.
    \label{eq:detVoltage}
\end{equation}

\begin{figure}
	\centering
	\includegraphics[width=\columnwidth]{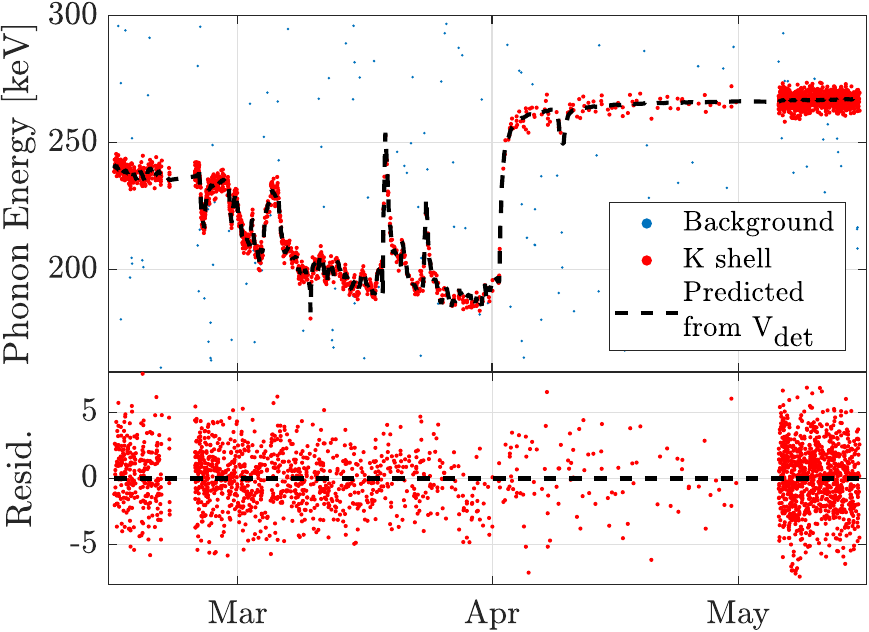}
	\caption{The drift in the total phonon energy for events in the 10.37 \kevee peak (from $^{71}$Ge $K$-shell decays) is well modeled by the measured variation of the detector voltage (\EQ\ref{eq:detVoltage}). Early in April the detector voltage stabilized at 75\,V. Three $^{252}$Cf calibrations were performed over the course of the run. The timing of the calibrations (Feb.\ 2--3, Feb.\ 20--23, and May 1--5), along with the 11.43 day half-life of $^{71}$Ge, is seen in the variable intensity of the $K$-shell decays. Data points labeled as background simply represent events originating from sources other than $K$-shell decays.}
	\label{fig:energyCorr}
\end{figure}

\noindent
We found that the parasitic resistance was correlated with the temperature of the room that housed the electronics. In April 2015 we adjusted the environmental conditions of this room to increase the parasitic resistance, thus lowering the leakage current  and stabilizing the detector voltage at 75\,V. Prior to April 2015, the detector voltage drifted between 50\,V and 75\,V. This resulted in $\sim$30\% variations in the total phonon energy scale, shown in \FIG\ref{fig:energyCorr}. We correct for this variation in the analysis, accounting for the small difference in the correction factor for nuclear versus electron recoils.

Following the stabilization of the detector voltage at $V_{\textrm{det}} = 75$\,V, the phonon noise performance worsened, indicating that the optimal operating voltage was slightly less than 75\,V. Based on these two distinct operating conditions---bias voltage stability and noise performance---we divided the \runThree data set into two periods:  \perOne and \perTwo (before and after April 1, respectively).  This division facilitates optimization of certain stages of the analysis, which were performed separately for the two periods.

\begin{figure}
	\centering
	\includegraphics[width=\columnwidth]{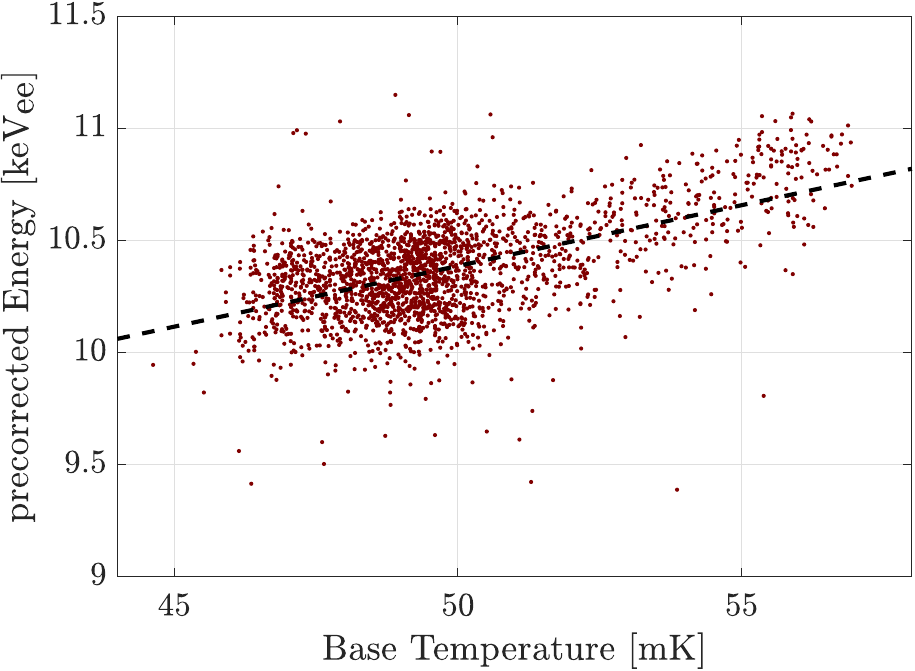}
	\caption{The reconstructed energies of the 10.37 \kevee peak events are positively correlated with the base temperature. This dependence is approximated as linear and corrected according to the fitted dashed line.}
	\label{fig:tempCorr}
\end{figure}

Additionally, the base temperature varied from 45 to 57\,mK over the course of the run. We applied a temperature-dependent empirical linear correction of up to $\sim$5\% to the energy scale. This correction was based on the positive correlation observed between the reconstructed energy of the $K$-shell events and the recorded base temperature, and is shown in \FIG \ref{fig:tempCorr}. 

After all corrections are applied, the energies of the $K$-, $L$-, and $M$-shell peaks agree with the expected values to within 3\%.

\subsection{Optimal Filter Energy and Position Reconstruction}
\label{sec:OF}

\begin{figure}
	\centering
    \includegraphics[width=\columnwidth]{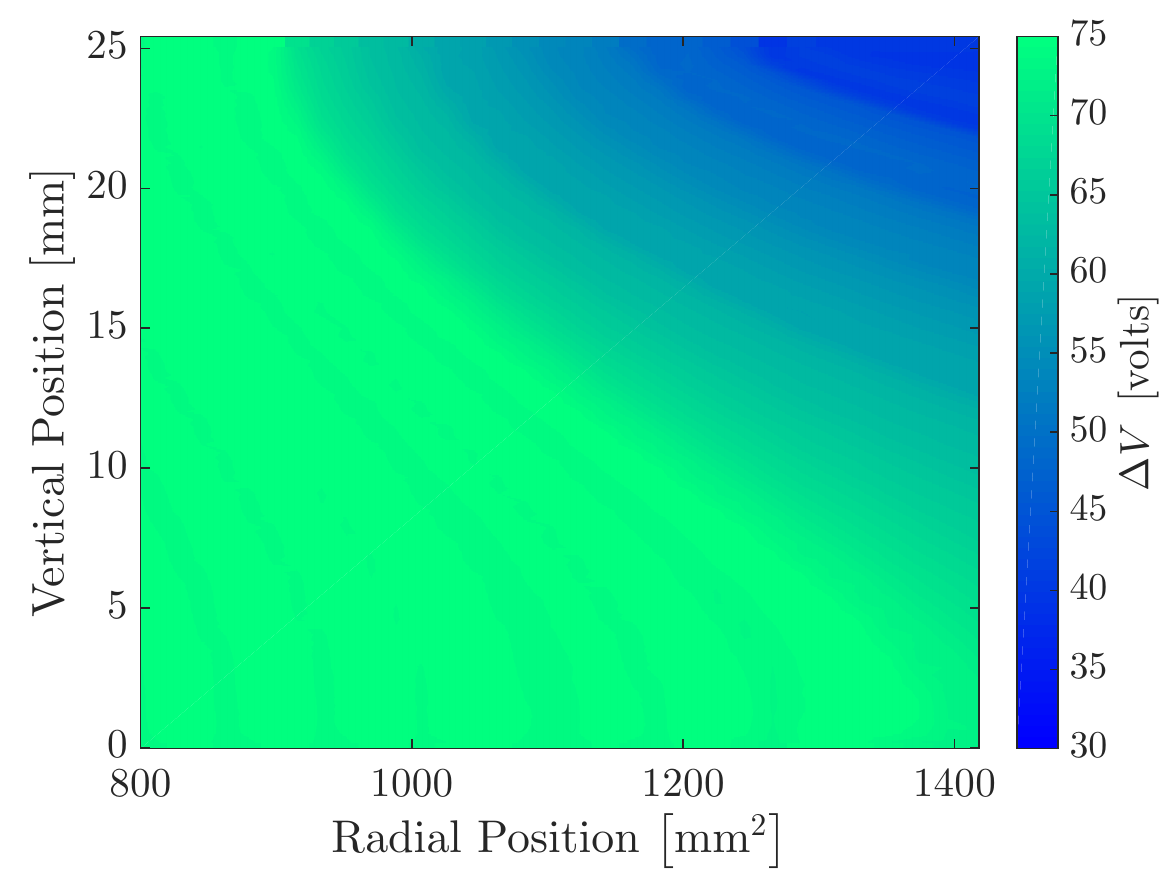}
    \caption{Calculated voltage map for high radius events, showing the difference in electric potential $\Delta V$ between the final collection points of the positive and negative charge carriers, as a function of initial position of the pair (plotted as radius squared vs. vertical position). Here, the top of the crystal is biased at +75\,V and the bottom is grounded. Charge carriers in the outermost (radius $> 800$~mm$^2$) detector annulus can experience less than the full detector bias voltage.}
    \label{fig:voltageMap}
\end{figure}

Because CDMSlite detectors have non-uniform electric fields, the NTL amplification and the reconstructed recoil energy vary with the location at which an event takes place inside the detector. For most events, $\Delta V$ in \EQ~\ref{eq:E_ntl} is equal to the full potential difference between the detector faces, resulting in maximal NTL amplification.  However, as shown in \FIG~\ref{fig:voltageMap}, near the detector sidewall $\Delta V$ can be smaller; the voltage drop experienced by an electron-hole pair (and thus the NTL amplification) can be reduced such that the reconstructed energy of some high-radius events is significantly lower.

While we cannot reconstruct the exact position of an event and thus correct for the specific reduced NTL amplification, we can calculate a parameter that correlates with the radial position of an event and use it to identify events at large radii.  We employ optimal filter algorithms \cite{And1979} to reconstruct the energy and position of events.  Optimal filters weight frequency components of the raw pulses to maximize the signal-to-noise ratio when fitting for the amplitude of a pulse, and the standard optimal filter algorithm assumes constant pulse shapes.

CDMSlite phonon pulses are slightly variable in shape, with differing proportions of ``slow'' and ``fast'' components from event to event. The former provides a measure of the total event energy, while the latter is sensitive to the event position---events occurring directly underneath a phonon channel cause a faster pulse rise time in that channel than in other channels.  We capture both types of information with a two-template optimal filter algorithm (2TOF)~\cite{Agnese2018}.  The first template is constructed from the average of many pulses, and then a second template is constructed from the average shape of the residual pulses (relative to the first template).  These correspond to the ``slow'' and ``fast'' templates, respectively.

The definition of the radial parameter, which we denote by $\xi$, remains the same as was used in \runTwo \cite{Agnese2018,Agnese2016}. It takes advantage of the phonon channel layout with an outer annulus and three inner wedge-shaped channels, comparing the amplitude of the fast template and the pulse start time between the outer and the inner channels.  The $\xi$ parameter identifies higher radius events that can experience reduced NTL gain and is used in \SEC\ref{sec:fiducialvolume} for fiducialization.

In addition to defining a radial parameter, the 2TOF is used to improve the event energy reconstruction. For each event, the best-fit amplitude from the fast template is used to apply a correction of up to $\sim$5\% to the leading order energy estimation, which is derived from the best-fit amplitude of the slow template using a separate optimal filter algorithm that specifically deweights the high-frequency components of the phonon pulses. We use the same correction procedure as that described in \SEC II.C of \REF\cite{Agnese2018}.

\subsection{Energy Resolution Model}
\label{sec:energyresolutionmodel}

We require a good model of the energy resolution in order to calculate the expected energy spectra for signal and backgrounds.
We model the total CDMSlite energy resolution as in \REF\cite{Agnese2018}:
\begin{align}
	\sigma_{\text{T}}(\Ereeeq) &= \sqrt{\sigma_{\text{E}}^2 + \sigma_{\text{F}}^2(\Ereeeq) + \sigma_{\text{PD}}^2(\Ereeeq)} \\
    						 &= \sqrt{\sigma_{\text{E}}^2 + B\Ereeeq + (A\Ereeeq)^2}.
	\label{eq:simpleRes}
\end{align}
\noindent
The energy-independent term $\sigma_{\text{E}}$ describes the baseline resolution and accounts for
electronics noise and any drift in the operating conditions. The Fano term $\sigma_{\text{F}}$ accounts for fluctuations in the number of generated charges~\cite{Fano1947} and is proportional to $\sqrt{\Ereeeq}$. The $\sigma_{\text{PD}}$ term
reflects the position dependence of the event within the detector due to the electric field, TES response, etc., and is proportional to $\Ereeeq$. Separating out the energy dependence we end up with the three model parameters $\sigma_{\text{E}}$, $B$, and $A$.

We use several measurements to determine the resolution model for \runThree.  We use randomly triggered events to determine the zero-energy noise distribution.  Additionally we use the widths of the $K$-, $L$-, and $M$-shell $^{71}$Ge activation peaks (see \SEC\ref{sec:expDescription:energyScale}) to determine the energy dependence of the resolution.
We fit these peaks with a combination of a Gaussian and linear background model in order to determine the width of the peaks.
\begin{table}
	\centering
	\begin{tabular}{@{} l l p{4.5} p{5.5} @{}}
		\hline
		Peak & & \multicolumn{1}{c}{Energy} & \multicolumn{1}{c}{Resolution} \\
		 & & \multicolumn{1}{c}{$\mu$ [\kevee]} & \multicolumn{1}{c}{$\sigma$ [\evee]} \\
		\hline
		$K$ shell & & 10.354,0.002 & 108,2.0 \\
		$L$ shell & & 1.328,0.003 & 36.3,2.0 \\
		$M$ shell & & 0.162,0.002 & 13.9,2.0 \\
		\multirow{2}{*}{Baseline} & \perOne & \multicolumn{1}{c}{0.0} & 9.87,0.04 \\
		 & \perTwo & \multicolumn{1}{c}{0.0} & 12.67,0.04 \\
		\hline
	\end{tabular}
	\caption{Reconstructed energies and resolutions of the $^{71}$Ge decay peaks and the baseline noise in CDMSlite \runThree.}
	\label{tab:run3resolutions}
\end{table}
\begin{figure}
	\centering
    \includegraphics[width=\columnwidth]{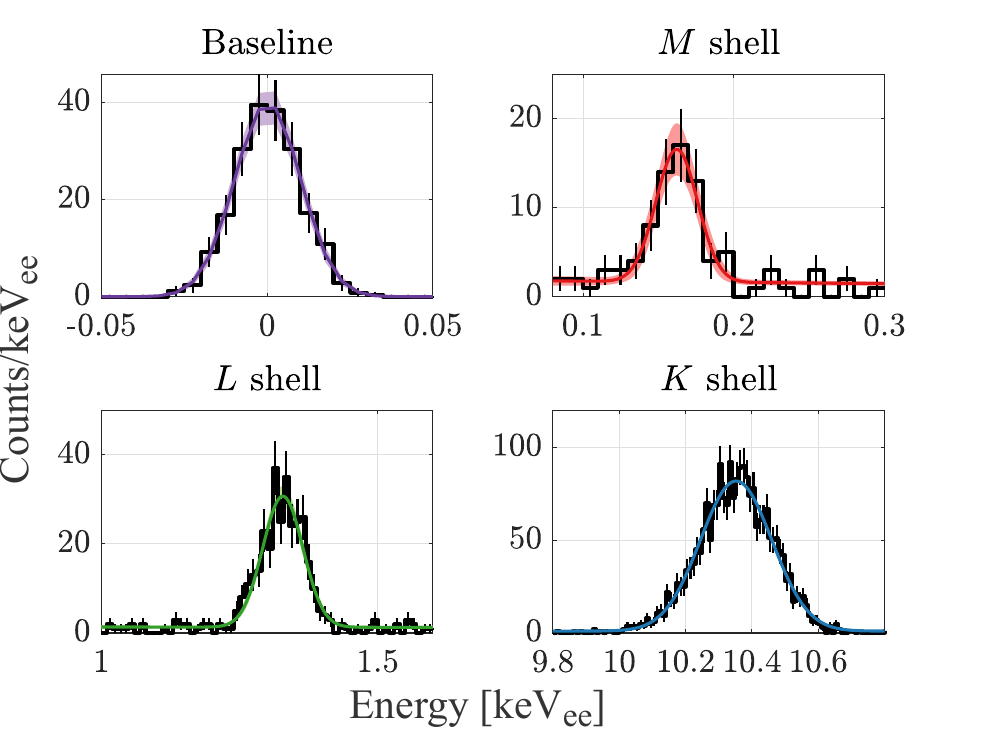}
    \caption{Fits of a Gaussian + linear background to the 
   energy spectra of zero-energy (baseline) events and events from each $^{71}$Ge activation peak.  The widths of the Gaussians are the energy resolution $\sigma$.}
    \label{fig:resPeakFits}
\end{figure}
\TAB\ref{tab:run3resolutions} gives the peak position $\mu$ and resolution $\sigma$ of each $^{71}$Ge peak, and \FIG\ref{fig:resPeakFits} shows the fits to the $K$-, $L$-, and $M$-shell peaks along with an example of the zero-energy noise distribution. 

Because the zero-energy baseline resolution varies with the applied bias voltage and with environmental conditions, all of which changed between \perOne and \perTwo, we calculate separate livetime-weighted average resolutions for each period.  These are given in \TAB\ref{tab:run3resolutions}.  The measured widths of the $K$-, $L$-, and $M$-shell peaks are consistent between \perOne and \perTwo, and so common values are used for both periods.

\begin{table}
	\centering
	\begin{tabular}{@{} l p{4.4} p{4.4} p{4.4} @{}}
		\hline
		 & \multicolumn{1}{c}{$\sigma_{\text{E}}$ [\evee]} & \multicolumn{1}{c}{$B$ [\evee]} & \multicolumn{1}{c}{$A$ $(\times 10^{3})$} \\
		\hline
		\perOne & 9.87,0.04 & 0.87,0.12 & 4.94,1.27 \\
		\perTwo & 12.7,0.04 & 0.80,0.12 & 5.49,1.13 \\
		\hline
	\end{tabular}
	\caption{Best-fit energy resolution parameters of the model in \EQ\ref{eq:simpleRes} for \perOne and \perTwo. 
    }
	\label{tab:run3resParam}
\end{table}

\TAB\ref{tab:run3resParam} gives the best-fit parameter values for the model of \EQ\ref{eq:simpleRes}. The coefficient $B$ is consistent between the periods, but is larger than the value predicted by measurements of the Fano factor, which is $B = 0.39$~\cite{Antman1966,Bilger1967}. The values of $A$ also agree within uncertainties between the two periods.

We apply this energy-dependent resolution model when calculating the expected energy distribution for the background and DM signal components.  We propagate uncertainties in the model parameters as systematic uncertainties in the profile likelihood fit of \SEC\ref{sec:likelihood}.

\section{Blinding Strategy}
\label{sec:salting}
\label{sec:saltingprocedure}

To avoid bias during the analysis, we adopted a blindness scheme to prevent analyzers from tuning the analysis to reach a desired result. Because instrumental noise is a significant and time-varying source of events, it is desirable to be able to see all events at each stage of the analysis. Therefore, rather than hiding events in the signal region, we implemented data ``salting'' in which a fraction of the DM-search events are replaced with artificial signal-like events.  This procedure effectively masks the true amount of DM signal in the data. The number and energy distribution of the artificial events were hidden from the analyzers. All analysis was done on the salted data until the last step, when we removed the added events, replaced the originals, and performed the final fits.  We opted to replace events with salt, rather than solely adding salt, to avoid the need to work around the sequential event IDs that are a feature of our data format. This had the added benefit of protecting against a possible tendency 
to overly tune cuts to the particular events in the salted data, since analyzers knew that some unknown number of events would be added back in after unblinding.

The salting procedure itself was openly developed and known to analyzers in advance, with a number of input parameters randomized and hidden until unblinding.  \TAB~\ref{tab:saltparams} lists these parameters, their allowed ranges (known in advance), and their randomly selected values that were hidden until unblinding.  The goal of this process was to produce a set of artificial events with an energy spectrum approximating a DM-induced nuclear-recoil distribution, with other event parameters (e.g. $\chi^2$ goodness-of-fit, radial parameter, etc.) consistent with detector-bulk events uniformly distributed in time and location. We generated artificial events using a pulse simulation similar to that described in \SEC VI.B of \REF\cite{Agnese2018} in which fast and slow pulse templates were added to in-run noise samples. The relative amplitudes of the two templates were determined by fitting each channel to calibration data. The salting procedure is described step-by-step below.

\paragraph{Select data events to replace:}
First, the number of events to replace with salt was selected randomly, and this number was kept hidden from the analyzers. The goal was to choose a number of events such that, after application of cuts, the remaining salt ``signal'' is between one and three times the predicted 90\% confidence level limit for the analysis. Based on the size of the CDMSlite Run 3 data set and the passage fraction of trial salt data sets generated with Run 2 data and cuts, the range was set to 280--840 events. Upon unblinding, the number of salt events was revealed to be 393. After application of the selection cuts described in \SEC\ref{sec:dataAnalysis} and \ref{sec:fiducialvolume} the number of salt events in the signal region was reduced to 105, which constituted 26\% of the number of true events that survived selection cuts in the signal region (401 events).

The events to be replaced by salt were chosen randomly from the data set with a uniform time distribution.  When events were replaced with salt, only the waveform data was changed, without changing any of the metadata such as trigger masks, timestamps, etc.  Therefore, only events that generated a trigger on the CDMSlite detector were considered. An additional preselection cut requiring the reconstructed energy to be greater than 3.5~keV in total phonon energy removed the majority of cryocooler-induced low-frequency noise events (discussed in \SEC \ref{sec:dataAnalysis:LFN}), which represent the largest source of non-uniformity in the event time distribution.  To select an event to replace, a random time was chosen within the CDMSlite Run 3 duration, weighted by the experiment livetime in one-day bins, and the nearest event passing preselection cuts was selected. If the chosen time was between data series, it was discarded. This process was repeated until the chosen number of events was selected.     

\paragraph{Choose an energy for each salt event:}
The event energies were chosen from an exponential distribution with a constant offset:

\begin{equation}
\label{eq:saltenergy}
P(E) \propto C + (1/D)\exp^{-E/D} ; \quad E \in [0.05, 5]~\text{\kevee},
\end{equation}

\noindent where the exponential component was chosen to roughly approximate a WIMP spectrum and the constant offset was chosen so that salt existed over the analysis energy region of interest. $C$ and $D$ are randomized hidden parameters, sampled logarithmically from $1/3$ to 3~keV$_\text{ee}$$^{-1}$  for $C$, and from $0.5$ to $2$~\kevee for $D$.  The chosen energy was also restricted from 0.05 to 5~\kevee to match the expected signal region of interest. The randomly selected parameters used were $C=0.6967$~keV$_\text{ee}$$^{-1}$ and $D=1.299$~\kevee, resulting in a nearly uniform distribution of salt events over the CDMSlite region of interest. 

\paragraph{Construct the artificial pulses:}
For each salt event, we constructed six artificial pulses (one for each phonon and charge channel). Each pulse, in turn, was constructed from the sum of a baseline noise waveform sampled during data acquisition, and the fast and slow templates used for 2TOF reconstruction.

The fast and slow pulse templates were scaled based on the reconstructed amplitudes of calibration events.  For each salt event, we randomly chose a calibration event near the target energy from the set of all calibration events passing basic preselection cuts.  These cuts included selection for good values for the bias voltage, current, base temperature, and the phonon pulse shape $\chi^2$ and noise event $\Delta\chi^2$ cuts described in \SEC\ref{sec:cutOverview}. Initially, we chose only from calibration events within 10~\evee\ of the target energy, after rescaling for corrections from varying parasitic resistance and temperature.  If no events were found in this window (excluding events that were already used for salt), the search was repeated with the range extended to 50 and then 100~\evee.   All reconstructed amplitudes were scaled by the ratio of the target salt energy to the calibration event energy, maintaining the relative amplitudes of the fast and slow templates.  In this way we produced salt events mimicking uniform bulk event distributions (e.g.\ in the radial parameter) without specifically modeling any of those variables. 

\paragraph{Pre-release validation:} 
Prior to beginning the salting procedure, a volunteer with substantial analysis experience was chosen to inspect the resulting salt. Several distributions were inspected with and without salt highlighted, to ensure that the salt did not significantly deviate from the data. When problems were identified, a fix was implemented, and the entire salting process was restarted. After validation, the pre-release inspector was excluded from any further analysis of the salted data set. 

\begin{table}
	\centering
	\begin{tabular}{@{} m{10em} p{4.4} p{4.4} p{4.4} @{}}
		\hline
		Parameter & \multicolumn{1}{c}{Range} & \multicolumn{1}{c}{Weight} & \multicolumn{1}{c}{Actual Value} \\
		\hline
        Number of salt events & \multicolumn{1}{c}{280--840}  &  \multicolumn{1}{c}{linear} &  \multicolumn{1}{c}{393} \\
		Spectrum constant weight ($C$ in Eq.~\ref{eq:saltenergy}) & \multicolumn{1}{c}{$1/3$--3} & \multicolumn{1}{c}{log} & \multicolumn{1}{c}{0.6967} \\
		Spectrum exponential slope ($D$ in Eq.~\ref{eq:saltenergy}) & \multicolumn{1}{c}{0.5--2} & \multicolumn{1}{c}{log} & \multicolumn{1}{c}{1.299} \\
		\hline
	\end{tabular}
	\caption{Randomized parameters used to generate the unknown salt data set. The units of the second and third row are \kevee and keV$_\text{ee}$$^{-1}$ respectively. The allowed range of parameters was known in advance, while the final value was hidden until unblinding after all cuts were finalized. For parameters with logarithmic weighting we randomly chose values from a uniform distribution for the logarithm of those parameters between their upper and lower limits.}
\label{tab:saltparams}
\end{table}

\section{Quality Cuts}
\label{sec:dataAnalysis}

A set of data quality cuts removes instrumental noise triggers, poorly reconstructed events, and periods when the detector was  behaving anomalously. Because this analysis employs profile likelihood methods to search for DM---fitting background and signal models to events that pass all cuts---it is imperative to identify and remove all instrumental noise events whose distributions cannot be modeled with a probability distribution in the fit. We use multivariate techniques in the lowest energy range of the analysis, where the experiment is most sensitive to DM particles with mass $<$\,10~\gev, to reduce instrumental noise leakage to less than 1 event while maintaining as low of an energy threshold as possible.

\subsection{Overview of Data Quality Cuts}
\label{sec:cutOverview}
We accept only events for which the power supply bias voltage was set to 75~V. We also remove any events in time coincidence with the NuMI neutrino beam~\cite{Michael2008}, including events whose time relative to the NuMI beam cannot be determined due to missing GPS information.

Cuts remove time intervals with easily identified anomalously high trigger rates. The ``prepulse,'' a $\sim$1~ms length of waveform data preceding the trigger and read out with each event pulse, is used to monitor noise and reject events with elevated noise.  Specifically we remove events in which the variance of the prepulse samples exceeds the average variance for events in the same three-hour data series by more than $4\sigma$.  We also designed cuts to remove electronics glitch events, which arise from instrumental noise and are characterized by pulse shapes with faster rise and fall times than signal pulses.\footnote{Throughout this section ``signal'' refers to good events caused by energy deposition in the detector, and ``background'' refers to instrumental noise events.} These cuts identify glitches that caused multiple detectors to trigger, glitches in the outer charge channel of the detector that could be coincident with phonon triggers, and glitches that are similar to signal events in all but pulse shape. Events with particularly bad noise in the charge waveforms were removed. Events that did not cause a trigger on the CDMSlite detector were also removed.  These cuts (excluding the pulse-shape glitch cut whose efficiency is considered separately) reduced the \runThree livetime from 66.9 to 60.9~days.

Due to their low interaction probability, DM particles are expected to interact at most once in our detector array. Therefore events that deposit energy above threshold in both the CDMSlite detector and a second detector are removed. The coincidence window used for identifying multidetector events was $-$200~$\mu$s to $+$100~$\mu$s around the CDMSlite detector's trigger time.  Events coincident with the muon veto surrounding the experiment are also removed, where a coincidence window of $-$185~$\mu$s to $+$20~$\mu$s around the event trigger time was used. These two cuts have a combined signal efficiency of 98.94\,$\pm$\,0.01\,\%.

Information from pulse-shape fits can discriminate signal events from instrumental noise events having a characteristic pulse shape.  Six different templates are fit to each event using the optimal filter method: a signal template, a square pulse template, an electronics glitch template with fast rise and fall times, and three low-frequency noise (LF noise) templates. The instrumental noise templates were created by identifying instrumental noise events in test processings of the data set, and averaging a collection of the raw pulses from the different instrumental noise sources. Three different LF noise templates were created because the LF noise assumes different pulse shapes, discussed in \SEC \ref{sec:dataAnalysis:LFN}.  

The $\chi^2$ values for each fit are used to classify and remove instrumental noise events. First, events with a high $\chi^2$ value for the signal-template fit are irregularly shaped (e.g. from event pileup or pulse saturation) and are removed. To remove particular classes of instrumental noise events, we use the difference of $\chi^2$ values between the different template fits: 
\begin{equation}
	\Delta \chi^2_{\text{LF,glitch,square}} \equiv \chi^2_{\text{OF}} - \chi^2_{\text{LF,glitch,square}},
\end{equation}
where OF corresponds to the standard signal-template fit, and LF, `glitch' and `square' correspond to the fits using the LF noise, glitch and square pulse templates respectively. Lower values of $\Delta \chi^2$ indicate events that have a more signal-like shape.

Glitch and square events have relatively uniform pulse shapes and do not resemble the signal pulse shape. Therefore, a single template for each is sufficient to efficiently discriminate against these event types. 

The $\Delta \chi^2_{\text{glitch}}$ and $\Delta \chi^2_{\text{square}}$ distributions for good signal events are parabolic as a function of event energy, and so we use simple two-dimensional cuts defined in the $\Delta \chi^2_{\text{glitch}}$ vs. energy and $\Delta \chi^2_{\text{square}}$ vs. energy planes. The signal efficiency of these cuts is energy dependent and $>$\,80\% down to the analysis threshold.

\subsection{Low-Frequency Noise Discrimination}
\label{sec:dataAnalysis:LFN}
Broadband low-frequency ($<$\,1~kHz) noise due to vibrational sources, shown to be primarily generated by the cryocooler that provides supplemental cooling power for the experiment, dominates the trigger rate for the CDMSlite detector~\cite{Agnese2018}.

In contrast to the other classes of instrumental noise events, LF noise events have variable pulse shapes and overlap substantially with the bandwidth of signal pulses. LF noise is therefore significantly more challenging to remove while maintaining high signal efficiency. Three LF noise templates are used to help identify a wide variety of LF noise shapes with $\Delta \chi^2_{\text{LF}}$ parameters.  Above reconstructed energies of $\sim$250 \evee, where the signal to noise in the waveforms is sufficiently high that LF noise events can be identified relatively simply, we use cuts on $\Delta \chi^2_{\text{LF}}$ values to remove LF noise events. Because discriminating against LF noise while maintaining signal efficiency is increasingly challenging at low energy, below $\sim$250 \evee we use boosted decision trees (BDTs) to improve the discrimination power of the LF noise cuts. In particular, we tune two BDT-based cuts using the bifurcated analysis technique \cite{Rusu2003,Nix2010} to ensure that less than one LF noise event leaks past the cuts.

\subsubsection{Bifurcated Analysis}

The bifurcated analysis method uses side band information (i.e. information outside of the signal region) to estimate a certain background's leakage past a set of quality cuts when no model exists for the background. We use this method for LF noise triggers because models for this background were found to be prone to significant systematic uncertainties.

The number of LF noise events leaking past a set of cuts is given by:
\begin{equation}
	N_{\text{leak}} = N_{\text{LF}} \cdot P(\text{cuts}),
\end{equation}
where $P(\text{cuts})$ is the passage fraction of the cuts and $N_{\textrm{LF}}$ is the number of LF noise events. While both $N_{\text{LF}}$ and $P(\text{cuts})$ are unknown, they can be estimated if there exist two uncorrelated sets of cuts that are both sensitive to LF noise events. Labeling the uncorrelated cuts A and B, denoting their known signal efficiencies as $\epsilon_A$ and $\epsilon_B$, denoting the unknown leakage fractions of LF noise events past the cuts as $L_A$ and $L_B$, and denoting the number of good (not LF noise) events as $N_{\textrm{G}}$, the numbers of events passing the individual and combined cuts are:

\begin{equation}
\begin{split}
\text{Pass Cut A : } & N_{AB} + N_{A\bar{B}} = \epsilon_{A} N_{\textrm{G}} + L_{A} N_{\textrm{LF}} \\
\text{Pass Cut B : } & N_{AB} + N_{\bar{A}B} =\epsilon_{B} N_{G} + L_{B} N_{\textrm{LF}} \\
\text{Pass Cut A\&B : } & N_{AB} = \epsilon_{A}\epsilon_{B} N_{\textrm{G}} + L_{A}L_{B} N_{\textrm{LF}}
\end{split}
\end{equation}
\noindent
where, for example, $N_{A\bar{B}}$ is the number of events that pass cut A but not cut B.

For uncorrelated cuts, the above system of equations can be solved to derive the number of LF noise events leaking past both cuts. For the case of cuts with 100$\%$ signal efficiency,
\begin{equation}
	N_{\text{leak}} = L_{A}L_{B} N_{\textrm{LF}} =\frac{N_{A\bar{B}}N_{\bar{A}B}}{N_{\bar{A}\bar{B}}},
    \label{eq:bifLeak}
\end{equation}
where side band information is used to estimate leakage into the signal region.
We include a small correction to \EQ\ref{eq:bifLeak} that accounts for the known $<$100$\%$ signal efficiencies of cuts A and B, where the signal efficiencies of the bifurcated cuts are measured using the method discussed in \SEC \ref{sec:sigEff:pulseshape}.

Two different LF noise cuts were designed that are uncorrelated so that the bifurcated analysis can be applied. These two cuts use three sets of parameters that are sensitive to LF noise in separate ways:
\begin{enumerate}
	\item the three $\Delta\chi^2_{\textrm{LF}}$ parameters from pulse shape fits to the three different LF noise templates.
	\item the $\hat{t}_{-}$ variable, which represents the time since the last cryocooler cycle. The cycle period is $\sim$0.8 seconds and LF noise causes triggers more frequently in the $\sim$0.2 seconds after the start of the cycle. This behavior is similar, though not identical, to that observed for the CDMSlite \runTwo detector \cite{Agnese2018}.
	\item the correlation of the phonon waveforms between the CDMSlite detector and the other detectors in the tower, because the vibrational sources producing LF noise triggers couple to all detectors in a tower.
\end{enumerate}
The first bifurcated cut (cut A) used primarily $\Delta\chi^2$ parameters to discriminate against LF noise; the second bifurcated cut (cut B) used primarily cryocooler time and cross-detector correlations. A BDT was used to reduce the bifurcated cuts to a single dimension (BDT A and BDT B). Both BDTs were trained using a `background' sample of LF noise (selected by removing events that fail the other quality cuts and removing events that are clearly good phonon pulses) and a `signal' sample of simulated good phonon pulses with noise. Details of the phonon pulse simulation are discussed in \SEC \ref{sec:sigEff:pulseshape}. For every event a BDT score is generated between $-$1 and 1, with a larger BDT score corresponding to a more signal-like event.

\begin{figure}
	\centering
    \includegraphics[width=\columnwidth]{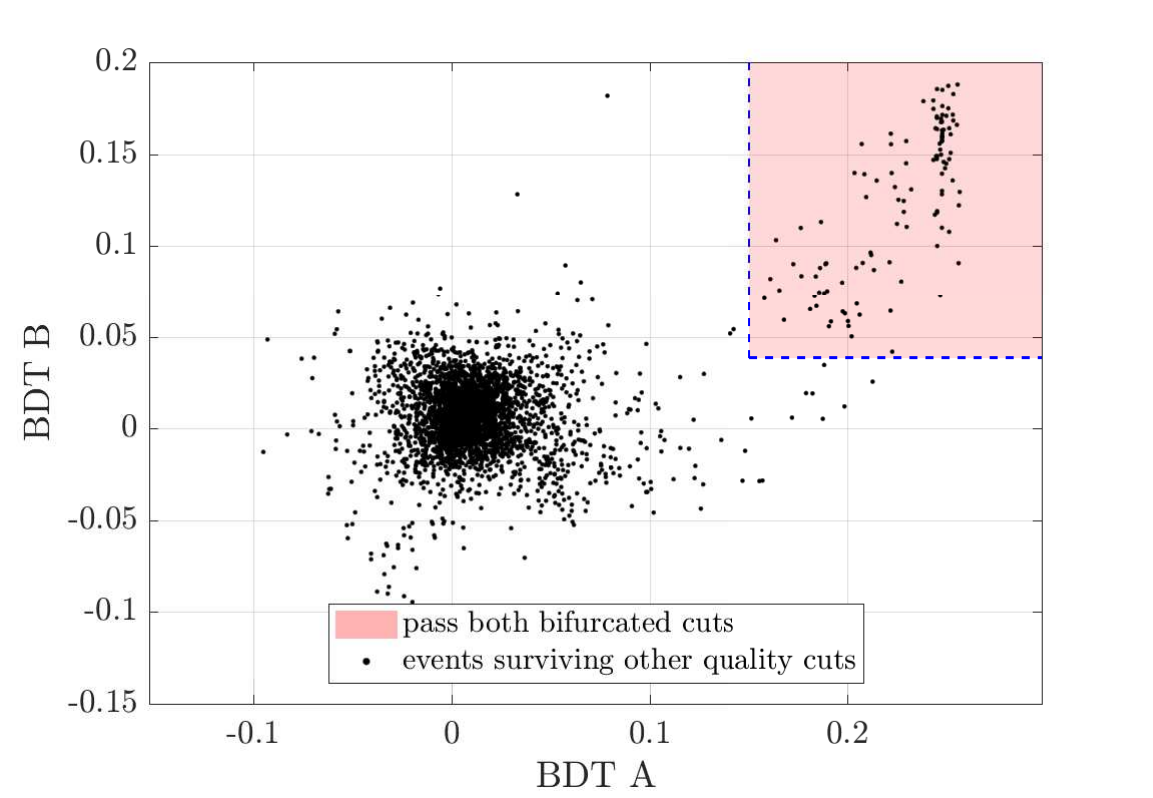}
	\caption{Two uncorrelated BDT variables are formed based on three sets of parameters that are sensitive to LF noise (see main text). The acceptance region of the bifurcated cuts using the BDT variables is shaded in the upper right.}
	\label{fig:bifurfactedPlane}
\end{figure}

\begin{figure}
	\centering
    \includegraphics[width=0.87\columnwidth]{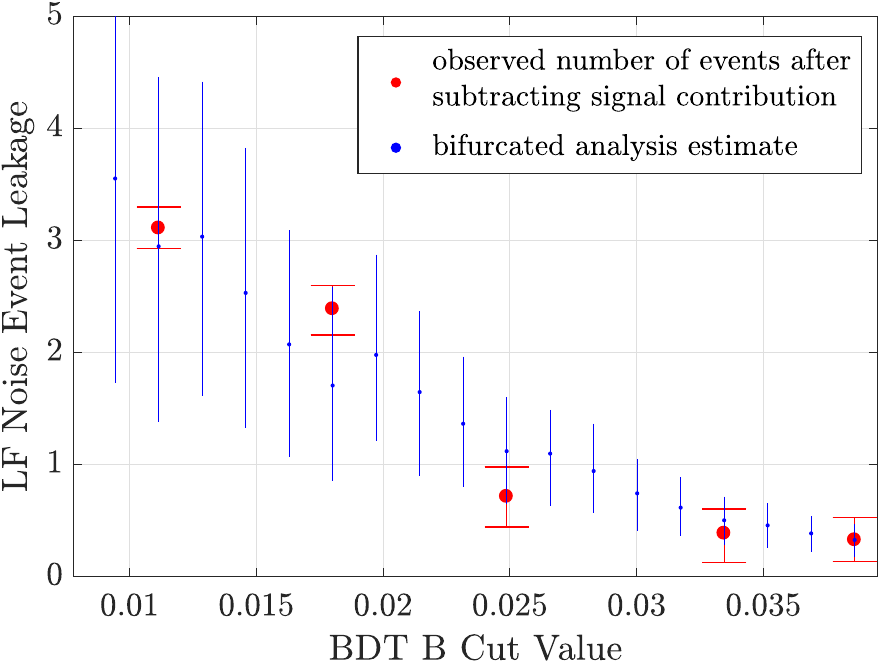}
	\caption{Variation of the number of background events leaking through the BDT cuts, as a function of the BDT B cut value.  The observed number of events, after subtracting the expected contribution from signal events, agrees with the bifurcated analysis estimate.}
	\label{fig:bifurfactedLoosen}
\end{figure}

The bifurcated analysis was then performed by placing cuts on the BDT A and BDT B scores and calculating the number of LF noise events leaking past the cuts, with cut values chosen such that the total LF noise event leakage is $<$\,1. Figure \ref{fig:bifurfactedPlane} shows the signal box (upper right) defined by the bifurcated cuts for \perOne; the estimated LF noise leakage for this period is $0.3 \pm 0.1$ events. A similar analysis on the \perTwo data gives an estimated event leakage of $0.1 \pm 0.1$ events. 

The choice of the cut location also assured minimal correlation between cuts. This was verified by the method of ``box relaxation.'' As a bifurcated cut is loosened, new events will enter into the signal box and the bifurcated leakage estimate will change. If the cuts are uncorrelated, the bifurcated analysis estimate will grow by the number of new events in the box (to within uncertainties). 
Because the BDT cut efficiency for signal events is not 100\% (see \SEC \ref{sec:sigEff:pulseshape}), 
we must correct for the contribution of signal events being added to the box as the cut is relaxed. We verified that the number of events entering the box matched the bifurcated analysis's prediction to within uncertainties, which is consistent with the cuts being uncorrelated and therefore supporting the validity of the leakage estimates. Figure~\ref{fig:bifurfactedLoosen} illustrates this agreement.

\section{Fiducial Volume Selection}
\label{sec:fiducialvolume}
A cut on the radial parameter $\xi$ (\SEC\ref{sec:OF}) defines a fiducial volume in order to remove events with reduced NTL gain (RNTLs) near the detector side wall.  The definition of this cut is improved compared to the CDMSlite \runTwo analysis~\cite{Agnese2016,Agnese2018}. 

We characterize RNTLs by modeling their distribution as a function of reconstructed energy and $\xi$.  The modeling is done in several steps: 

\begin{enumerate}[label=\Alph*.]
\item Determine the energy response of the detector, using the NTL gain as a function of position inside the detector (\SEC\ref{sec:energyRNTLs});
\item Determine the rates of events that contribute RNTLs into the signal region below 2 \kevee  (\SEC\ref{sec:SourceRNTL});
\item Model the distribution of RNTLs in $\xi$  (\SEC\ref{sec:2TRPRNTL});
\item Model the resolution of $\xi$ as a function of energy and $\xi$ (\SEC\ref{sec:radres});
\item Construct a Monte Carlo simulation based on these models to determine the expected distribution of RNTLs in the energy-$\xi$ plane, and define a cut in this plane to remove these events (\SEC\ref{sec:radMC}); and
\item Extend the cut above 2 \kevee where $\xi$ begins to change due to phonon-sensor saturation (\SEC\ref{sec:rc2kev}).
\end{enumerate}

\subsection{Energy Distribution of RNTLs}
\label{sec:energyRNTLs}
We use a smoothed histogram of the effective potential distribution shown in \FIG\ref{fig:voltageMap} to determine the energy response of the detector to a homogeneously distributed mono-energetic source of events.

We define the RNTLs to include any event whose recoil location results in a reconstructed recoil energy that differs from the true recoil energy by more than the $1\sigma$ detector energy resolution. This corresponds to events that see less than 93.3\% of the full bias voltage $V_{\textrm{det}}$.  For electron recoils, the measured event energy is reduced from the nominal expectation according to
\begin{equation}
E^{\mathrm{measured}} =  E^{\mathrm{nominal}}\times\frac{1+\frac{\Delta V}{\epsilon_\gamma}}{1+\frac{ V_{\textrm{det}}}{\epsilon_\gamma}},
\end{equation}
where $\Delta V$ is the potential difference experienced by charge carriers produced at the recoil location, and $V_{\textrm{det}}$ is the nominal potential difference.  

The shape of the voltage distribution is a source of systematic uncertainty for the distribution of RNTLs, and to account for this we perform the same analysis with an alternate voltage distribution containing more features in the voltage spectrum from the simulation. This predicts a slightly higher leakage rate of RNTLs given the same radial cut, and gives us a handle on the systematic uncertainty on the rate of RNTLs we expect to pass our radial cut.

\subsection{Identification of RNTLs}
\label{sec:SourceRNTL}
Following the analysis done in~\runTwo ~\cite{Agnese2018}, we use the 11.43~day half-life of the $^{71}$Ge produced during neutron calibrations to statistically differentiate $K$-shell capture events from other backgrounds in different regions of the energy-$\xi$ plane.  This study of $^{71}$Ge decay events finds that  $86 \pm 1 \%$ of events receive full Luke gain and thus are reconstructed at the correct energy, making the remaining $14\%$ RNTLs.   

We then use this fraction to calculate the number of RNTLs in our data set.  We first determine the number of $K$-shell events from $^{71}$Ge decays by fitting the time distribution of events in the $K$-shell line at 10.37~\kevee with a component that decays with the 11.43~day half-life of $^{71}$Ge plus a flat component due to other backgrounds.  

Using the known number and energy of the $K$-shell events and the shape of the tail in the $\Delta V$ distribution from the voltage map, we can then determine the expected number of $K$-shell RNTLs in our signal region below 2~\kevee.  The contributions of $L$-shell and $M$-shell events are estimated by scaling the number of RNTL events from $K$-shell decays by the theoretical branching ratios between these shells (see Table \ref{tab:cosmogenicEC}).  All RNTL events from $L$-shell or $M$-shell decays are in the energy region of interest for this analysis.

The rates of other backgrounds below the $L$ shell are determined by assuming that they are distributed uniformly in volume, energy, and rate, measuring their rate in a region free from RNTLs and $^{71}$Ge events ($\xi<-2\times 10^{-5}$, $E$~$\in[0.6,1]$~\kevee) and extrapolating to the full volume and energy range.  Similarly, the rate of events above the $L$ shell is extrapolated from a region higher in energy than the $L$ shell that is free from RNTLs and $^{71}$Ge events ($\xi<-2\times 10^{-5}$, $E$~$\in~[1.5,2]$~\kevee).

The final step for calculating the rate of RNTLs is to scale the rates by the energy-dependent efficiencies of all the other cuts.  We estimate that there are $133.1\pm7.6$ RNTLs in the signal region before applying a fiducial volume cut.

\subsection{Distribution of RNTLs in $\xi$}
\label{sec:2TRPRNTL}
The majority of RNTLs are measured only slightly lower in energy than their true energy, because the distribution of $\Delta V$ inside the detector peaks strongly at the nominal voltage.  Thus the energy regions just below the strong $K$ and $L$-shell $^{71}$Ge-decay peaks provide good samples of RNTLs whose properties can be studied to determine their distribution in the radial parameter $\xi$.

We model the radial distribution of RNTLs by defining a region in the radial parameter ($\xi\in[-2\times 10^{-5},+4\times 10^{-5}] $) outside of which we observe no RNTLs, and selecting events in this region within a small energy range below the $L$-shell capture peak (0.7--1.2 \kevee).  Creating a cumulative distribution function in $\xi$ for these events gives us an idea of the distribution of RNTLs in $\xi$. A systematic uncertainty on this distribution is estimated by removing the upper bound in $\xi$ while narrowing the energy window, which creates a distribution that predicts slightly more RNTLs passing the same cut.

\subsection{Resolution of $\xi$}
\label{sec:radres}
\begin{figure}
	\centering
    \includegraphics[width=\columnwidth]{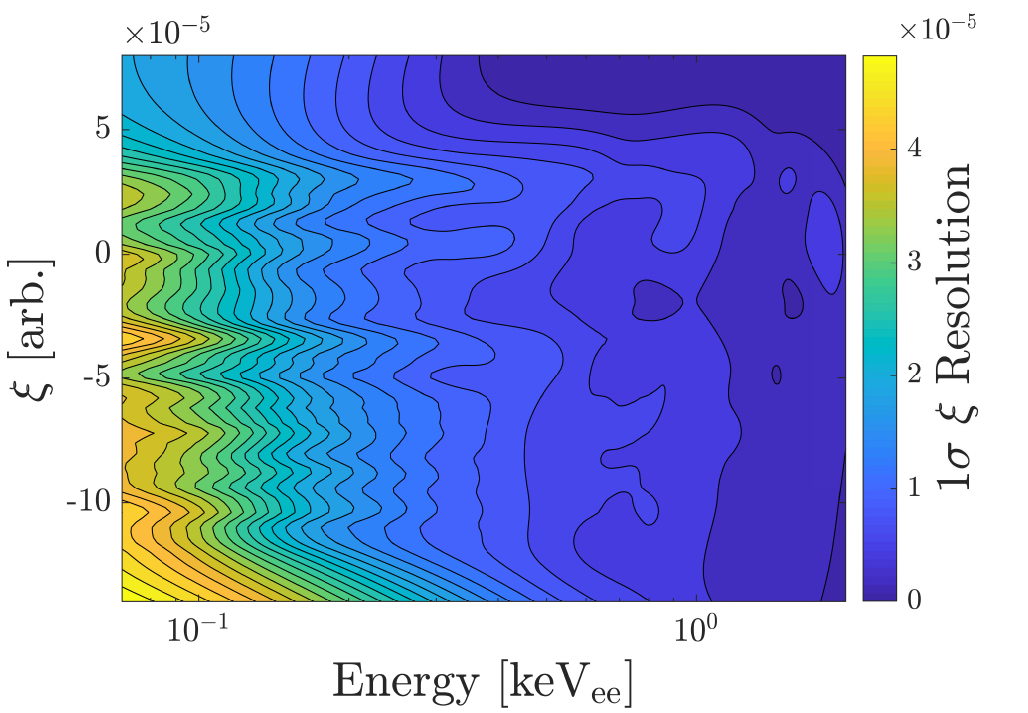}
	\caption{Resolution ($1\sigma$) for $\xi$ (radial parameter) shown as a function of $\xi$ and energy. At lower energy, the resolution worsens as the increased noise affects the reconstruction of the radial parameter.}
	\label{fig:radres}
\end{figure}

To model the resolution of $\xi$, we create sets of simulated events based on the 2TOF templates and fits of $^{71}$Ge $L$-shell capture events (a large sample that well represents the true $\xi$ distribution through the full range of $\xi$) in the manner done in Ref.~\cite{Agnese2018}.  Differently from what was done for the \runTwo analysis, we simulate each event with 100 different noise traces and use the resulting output to find the spread in $\xi$ due to the noise, as a function of $\xi$ and energy.  By fitting a Gaussian distribution to these sets of simulations, we build a model of the $\xi$ resolution as a function of ``true'' $\xi$ and energy (\FIG\ref{fig:radres}).

\subsection{RNTL Monte Carlo Simulation}
\label{sec:radMC}
Combining the expected energy distributions of RNTL events, the voltage map model, and the resolution model for $\xi$ as a function of energy, we can model the RNTL distribution throughout the full energy-$\xi$ plane.  We use a Monte Carlo method to sample these distributions and thus produce a prediction for the 2D probability distribution of the data in these variables.  We set a cut on $\xi$ as a function of energy based on this distribution, such that we expect $0.13 \pm 0.10_{\textrm{stat}} \pm 0.44_{\textrm{sys}}$ RNTLs passing the cut. The systematic error is estimated from Monte Carlo simulations with the alternate radial and voltage models (with the radial distribution of RNTLs being the larger contributor). The cut boundary was chosen such that the expected distribution of RNTLs passing the cut is uniform in energy between 0.07 and 2~\kevee. The radial parameter cut imposes an analysis threshold of 
70~\evee, which is determined by the lowest well-determined bound of the radial resolution model.

\subsection{Radial Cut above 2 keVee}
\label{sec:rc2kev}
\begin{figure}
	\centering
    \includegraphics[width=\columnwidth]{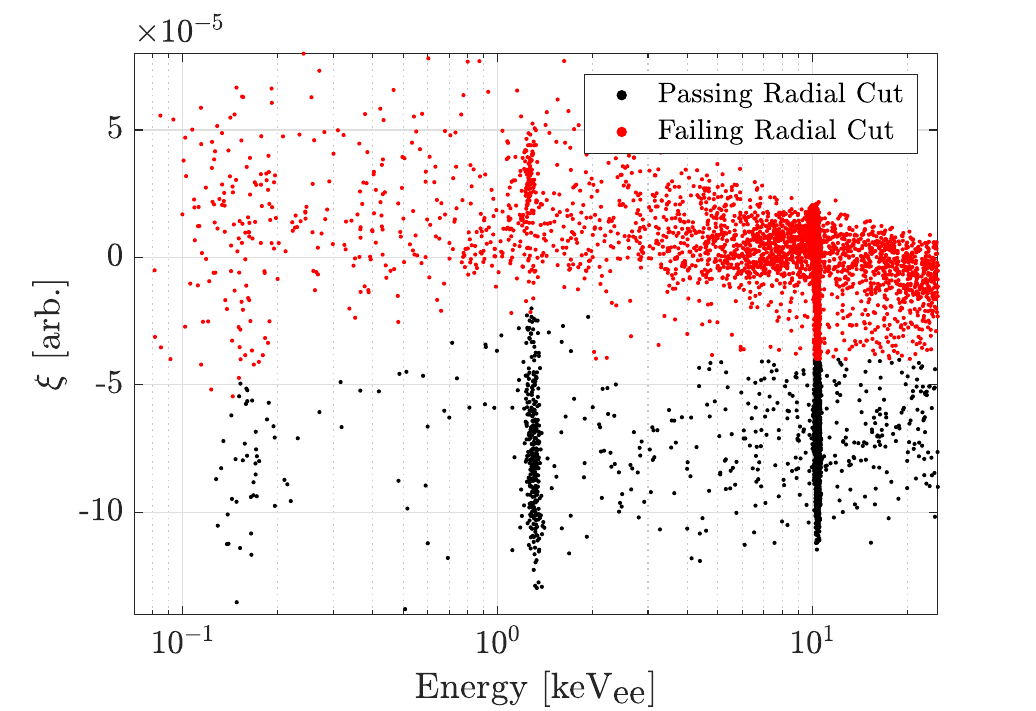}
	\caption{Distribution of the radial parameter $\xi$ vs. energy in the DM search data.  An energy-dependent cut on $\xi$ defines the fiducial volume below 2\kevee, while a stricter constant cut is used above 2\kevee. }
	\label{fig:radcut}
\end{figure}

In \SEC\ref{sec:sensitivity} we will estimate the sensitivity of the experiment to DM interactions based on background expectations derived from higher energies (5--25~\kevee) that are then extrapolated down into the signal region (0.07--2.0~\kevee).  Therefore, fiducialization at higher energies is needed. Above 2~\kevee, the threshold of the radial cut is set differently because we cannot model $\xi$ as well, due to saturation effects in the phonon pulse shape affecting the measured $\xi$.  Instead, we set a restrictive cut at $-4\times 10^{-5}$ in $\xi$ above 2~\kevee so that we expect zero RNTLs in this region.  The full range of data with the radial cut applied is shown in \FIG\ref{fig:radcut}.

\section{Signal Efficiency}
\label{sec:sigEff}

We calculate the DM signal efficiency of the \SEC\ref{sec:dataAnalysis} quality cuts and \SEC\ref{sec:fiducialvolume} fiducial volume cut by simulating raw pulses, processing them through the analysis pipeline to calculate the different cut variables, applying the cuts to them, and calculating their passage fraction as a function of energy.

The trigger efficiency is calculated with an alternate technique using information derived from events triggered on the rest of the detector array. It turns out that the trigger efficiency is a minor contributor to the total efficiency because the data quality and fiducial volume cuts are less efficient than the trigger at low energy.

\subsection{Data Quality Cuts}
\label{sec:sigEff:pulseshape}

The efficiency of the signal template $\chi^2$ cut, the $\Delta \chi^2_{\text{glitch}}$ and $\Delta \chi^2_{\text{square}}$ cuts, and the two BDT-based LF noise cuts is calculated using simulations of the total phonon pulse (i.e.\ the sum of the pulses for all phonon channels read out from the CDMSlite detector). These simulations depend on accurately representing the phonon readout noise in the pulses as well as the shapes of the true phonon pulses. We accomplish this by combining noise traces from randomly triggered events with noiseless phonon pulse templates. The true phonon pulses contain pulse-shape variations; to recreate these variations we use a linear combination of the fast and slow templates (see \SEC\ref{sec:OF}): $P = N \times (T_{s} + r T_f$). For each simulated pulse, we select values for the simulated pulse amplitude $N$ and the fast template component $r$ from a two-dimensional distribution of these parameters drawn from the full DM-search data set.  These simulated pulses span the energy range of interest for the analysis.

The cryocooler timing variable  $\hat{t}_{-}$ and waveform correlations between detectors are also recreated for the simulated pulses, which are inputs to the BDT-based LF noise cuts. The noise traces from which the simulated pulses are formed are uniformly distributed in $\hat{t}_{-}$. Because DM signal events should also be uniformly distributed in this variable, the simulated pulse uses the $\hat{t}_{-}$ variable from the noise trace. The noise traces also provide the detector-detector correlation variables. When the noise trace is acquired, the waveforms on the other detectors in the tower are also recorded. After adding the simulated phonon pulse $P$ to the noise trace on the CDMSlite detector, we calculate the waveform correlations between detectors. Finally, we calculate the BDT scores for the simulated data and apply the cuts. The combined efficiency of all data quality cuts, including the energy-independent multiples and muon veto cuts, is shown in \FIG \ref{fig:tieredEff}.

\subsection{Fiducial Volume}
\label{sec:sigEff:fiducial}

The efficiency of the fiducial volume cut can be measured with techniques similar to those used to construct the RNTL model in \SEC\ref{sec:fiducialvolume}.  We use a Monte Carlo simulation based upon the resolution model of $\xi$ to simulate the radial parameter distribution for events having the full NTL amplification.  We model the $\xi$ distribution for these events after that of events with reconstructed energies in the $L$-shell line.  We statistically subtract the small contribution of non-$^{71}$Ge backgrounds from this distribution and deconvolve the radial-parameter resolution at 1.3 \kevee.  

The result is what is expected to be the underlying ``true'' distribution of $\xi$ for events at the $L$-shell energy.  We then use the model of $\xi$ to scale this distribution according to energy, thereby creating energy-dependent probability distributions for $\xi$.  Finally, we apply the radial cut to these simulated distributions, and by doing so obtain the efficiency of the fiducial volume cut for events with full NTL amplification.  

To obtain the full efficiency of the radial cut, this number must be multiplied by the percentage of events reconstructed at the correct energy (i.e. having the full NTL amplification), as the resolution model for $\xi$ is valid only for those events at the correct energy.  We specifically set the cut to remove all RNTLs; we therefore estimate the full efficiency of the radial cut by multiplying by the percentage of non-RNTLs (86$\%$).

\subsection{Trigger Efficiency}
\begin{figure}
	\centering
	\includegraphics[width=\columnwidth]{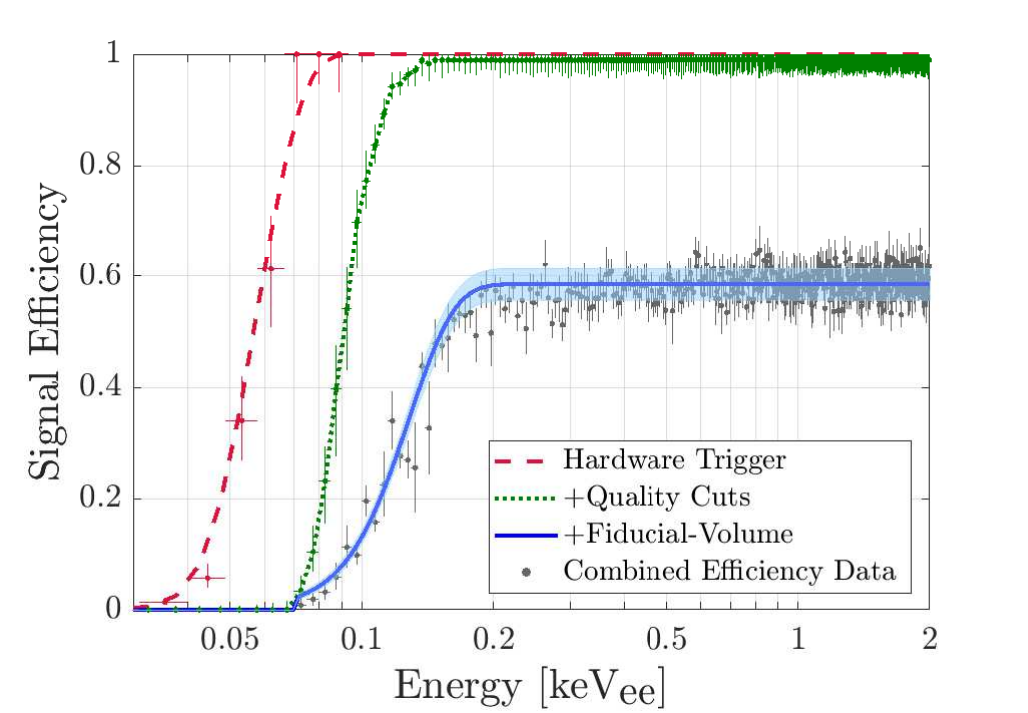}
	\caption{The signal efficiency with successive application of the trigger efficiency, quality cuts efficiency, and fiducial volume cut efficiency.   The final data is included with statistical and systematic 1$\sigma$ uncertainty. Fitting the efficiency model to these data gives the final (blue) efficiency curve and the corresponding $\pm$\,1$\sigma$ uncertainty band.} 
	\label{fig:tieredEff}
\end{figure}

The data acquisition system for CDMSlite issues a trigger and reads out events only when an energy deposition is large enough to create a significant increase of the signal above the baseline noise and thus exceed the trigger threshold. To measure the trigger efficiency we select events that have triggered in the other active detectors because they are an unbiased sample of events with respect to the CDMSlite detector's trigger. The trigger efficiency is then given by the fraction of events at any given energy (measured in the CDMSlite detector) that also generate a trigger in the CDMSlite detector. We use \cf calibration data, which has a significantly higher event rate than the DM-search data, to decrease the statistical uncertainty of the trigger efficiency measurement.  To model the trigger efficiency as a function of energy, we  fit an error function  to the data using the same method as was used in the \runTwo analysis \cite{Agnese2018}.  The final trigger efficiency curve is shown in \FIG \ref{fig:tieredEff}. Above 0.09~\kevee the trigger efficiency is equal to 100\% with negligible statistical uncertainty.

\subsection{Parametrization}
\label{sec:sigEffParam}

The efficiencies for the trigger, the data quality cuts, and the fiducial volume cut are combined by multiplying their mean values and propagating their respective uncertainties.

Incorporating the signal efficiency into the likelihood, described in \SEC \ref{sec:likelihood}, is most easily accomplished by parameterizing the final efficiency using a functional form with a limited number of model parameters. We find that a three-parameter error function,

\begin{equation}
\label{eq:efficiencyerf}
	h \left(E; \vec{\mu}_{e} \right) = \mu_{e_1} \times \left[1 + \text{erf}\left( \frac{E-\mu_{e_2}}{\sqrt{2} \mu_{e_3}} \right)\right],
\end{equation}

\noindent
is a good parametrization of the total efficiency curve. This simple efficiency parametrization deviates from the data slightly ($\lesssim4$\,\%) in the 0.15--0.4~keV$_{\text{ee}}$ range. We verified that this deviation results in a negligible change in the expected DM sensitivity. 
We determine the best-fit values of $\mu_{e_1}$, $\mu_{e_2}$, and $\mu_{e_3}$ as well as the covariance between these parameters, denoted by a matrix $\textbf{E}$. This matrix is used to propagate uncertainties in the efficiency parameters into the profile likelihood fit of \SEC \ref{sec:likelihood}.

Because the radial cut imposes an analysis threshold cutoff at 70~\evee, as described in \SEC \ref{sec:fiducialvolume}, we set the efficiency below this energy to zero, as seen in \FIG \ref{fig:tieredEff}.

\section{Background Models}
\label{sec:backgrounds}

The \scdms Soudan experiment was located at the Soudan Underground Laboratory with 2090 meters water equivalent overburden. The cryostat was surrounded by layers of shielding that blocked almost all external radiation, such as $\gamma$-rays and neutrons from the cavern walls. Thus, the radioactivity of the shielding and the other apparatus materials was the dominant source of background. We use Monte Carlo simulations, as well as data-driven fits, to model these backgrounds.

The backgrounds modeled for this analysis are as follows: cosmogenic activation of the crystal, specifically tritium, $^{68}$Ga, $^{65}$Zn, and $^{55}$Fe; neutron activation from \cf calibration; Compton scattering of gamma rays emitted from primordial isotopes in the apparatus materials; and $^{210}$Pb contamination on the surfaces of the detector and its copper housing.

\subsection{Cosmogenic Activation}
Cosmic rays can cause spallation resulting in cosmogenic activation of the crystals and apparatus materials during fabrication, storage, and transportation above ground. In germanium detectors, tritium contamination is a significant background, with contributions from  other isotopes that decay primarily either by $\beta$-decay or electron capture (EC). The additional cosmogenically-produced isotopes that undergo $\beta$-decay have endpoints of $\mathcal{O}$(MeV) and relatively small production rates. These can generally be ignored. The isotopes that undergo EC give discrete lines in the detectors below $\sim$10~keV and were observed in the CDMSlite \runTwo spectrum~\cite{Agnese2018b}. We describe analytic models for the tritium beta-decay spectrum and the EC lines.

\subsubsection{Tritium}
Non-relativistic $\beta$-decay theory suffices to model tritium's decay spectrum because its endpoint, or $Q$-value, satisfies the relationship $Q \ll m_e c^2$, where $m_e$ is the electron mass. The  distribution of the electron's kinetic energy $E_{\text{KE}}$ is described by 
\begin{equation}
	\begin{split}
		f_{\mathrm{tritium}}(E_{\text{KE}}) =& C \sqrt{E_{\text{KE}}^2 + 2E_{\text{KE}} m_e c^2}\left(Q - E_{\text{KE}}\right)^2 \\
        &\times\left(E_{\text{KE}} + m_e c^2\right) F\left(Z, E_{\text{KE}}\right),
		\label{eq:tritiumBeta}
	\end{split}
\end{equation}
where $C$ is a normalization constant and $F\left(Z,E_{\text{KE}}\right)$ is the Fermi function~\cite{Krane1988}. The non-relativistic approximation for the Fermi function is given by
\begin{equation}
	F\left(Z,E_{\text{KE}}\right) = \frac{2\pi\eta}{1 - e^{-2\pi\eta}},~\text{with}~ \eta = \frac{\alpha Z (E_\text{KE}+m_e c^2)}{p c}.
	\label{eq:tritiumFermi}
\end{equation}
Here $Z$ is the atomic number of the daughter nucleus, $\alpha$ is the fine structure constant, and $p$ is the electron's momentum~\cite{Povh2008}. The analytical description given by \EQS\ref{eq:tritiumBeta} and~\ref{eq:tritiumFermi} describes the tritium background used for the likelihood analysis. 

\subsubsection{Electron Capture Peaks}
\begin{table}
	\centering
	\begin{tabular}{@{} l d{3.3} c d{3.3} d{1.5} d{3.3} d{1.5} @{}}
		\hline Shell: 
		 & \multicolumn{2}{c}{$K$} & \multicolumn{2}{c}{$L_1$} & \multicolumn{2}{c}{$M_1$} \\
		 & \multicolumn{1}{c}{$\mu$ } & \multicolumn{1}{c}{$\Lambda$} & \multicolumn{1}{c}{$\mu$ } & \multicolumn{1}{c}{$\Lambda$} & \multicolumn{1}{c}{$\mu$} & \multicolumn{1}{c}{$\Lambda$} \\
		\hline
		$^{68}$Ge/$^{71}$Ge & 10,37 & 1.0 & 1,30 & 0,1202 & 0,160 & 0,0203 \\
		$^{68}$Ga & 9,66 & 1.0 & 1,20 & 0,1107 & 0,140 & 0,0183 \\
		$^{65}$Zn & 8,98 & 1.0 & 1,10 & 0,1168 & 0,122 & 0,0192 \\
		$^{55}$Fe & 6,54 & 1.0 & 0,77 & 0,1111 & 0,082 & 0,0178 \\
		\hline
	\end{tabular}
	\caption{Cosmogenic isotopes that decay via electron capture and are present in the measured CDMSlite spectrum. The shell energies $\mu$, given in keV, are from \REF\cite{Thompson2009}. The amplitudes $\Lambda$, from \REF\cite{Schonfeld1998}, are normalized with respect to the $K$~shell.}
	\label{tab:cosmogenicEC}
\end{table}
The cosmogenic isotopes that decay via EC and are present in the measured CDMSlite spectrum are listed in \TAB\ref{tab:cosmogenicEC} with their shell energies and relative amplitudes, normalized to the $K$ shell. The observed energy distribution is a Gaussian peak at the energy of the respective shell with a width given by the detector's energy resolution.

In our background model, the amplitude ratio between the $K$-, $L$- and $M$-shell peaks is assumed to be as given in \TAB\ref{tab:cosmogenicEC}. The contribution of each EC isotope to the spectrum is given by an equation of the type

\begin{equation}
	f_{\text{ECpeaks}}(E) = \sum_{i = K,L,M} \frac{\Lambda_i}{\sigma_i \sqrt{2\pi}} ~\text{exp}\left[-\frac{1}{2}\left(\frac{E - \mu_i}{\sigma_i}\right)^2 \right],
	\label{eq:ECpeaks}
\end{equation}
where $\Lambda_i$ are the amplitudes of the respective shells, $\mu_i$ are the shell energies, and $\sigma_i$ are the energy resolutions at the respective energies.

By modeling the EC peaks with \EQ\ref{eq:ECpeaks}, the number of events in the $K$ shell is the only free parameter in the likelihood fit, with the other peak amplitudes determined from the branching ratios.

\subsection{Electron Capture of $^{71}$Ge}
Neutrons from the $^{252}$Cf calibration source can be captured by $^{70}$Ge, creating $^{71}$Ge, which undergoes EC.  Although we use these peaks for calibration (see \SEC\ref{sec:expDescription:energyScale}) and although they decay with a  half-life 11.43~days, they are still a source of background. They are modeled using the same functional form as the cosmogenic EC peaks (\EQ\ref{eq:ECpeaks}) with the one exception that due to the large overall number of events the L$_2$ peak is not negligible and is thus included in the fit. This component, omitted from Table~\ref{tab:cosmogenicEC}, has an energy of 1.14~keV and relative amplitude of 0.0011.

%

\subsection{Compton Scattering}
The Monash University Compton Model~\cite{Brown2014} calculates properties of the scattered incident photon and the detector's recoiling electron by accounting for the atomic binding energy.
This treatment is necessary to replicate the phenomenon of ``Compton steps''---step-like features created in the energy spectrum because the detector collects at least the binding energy of any freed electron. For example, the electrons in the $K$~shell of germanium have a binding energy of 11.1~keV. This energy is deposited in the detector due to the reorganization of the electron shells, along with any additional energy that is given to the freed electron by the incident gamma. Thus, an electron from the $K$~shell can never deposit less than 11.1~keV in the detector, and likewise for electrons in the other atomic shells. Na\"{i}vely we would expect the number of electrons in each shell to determine the relative size of the steps; however details of the electron wave functions can also affect the step size. The Compton steps have been directly observed in silicon detectors~\cite{Ramanathan2017}. In germanium, only the $K$-shell step has been measured directly, and so other methods must be used to estimate the lower energy steps~\cite{Barker2017}.

The dominant contributors to the Compton background are the radiogenic photons from trace amounts of contamination in the experimental materials. These originate from the shield materials (polyethylene and lead) as well as the cryostat and towers (copper). To estimate the shape of this particular background, we carried out a \geant simulation~\cite{Agostinelli2003,*Allison2006,*Allison2016} of $^{238}$U decays originating from the cryostat cans. The spectrum of deposited energy in the CDMSlite detector from this decay was determined to be characteristic of all bulk contamination. We fit a model consisting of a sum of error functions, 
\begin{equation}
	f_C(E) = \Lambda_0 +  \sum_{\substack{i = K,L,\\M,N}} 0.5 \Lambda_i \left( 1 + \text{erf}\left[\frac{E - \mu_i}{\sqrt{2}\sigma_i}\right]\right),
	\label{eq:comptonModelFinal}
\end{equation}
to the the simulated events that scatter once in the CDMSlite detector. The location of each step is given by $\mu_i$, 
while $\sigma_i$ is the energy resolution at that energy given by the energy resolution model of \SEC\ref{sec:expDescription}.  The $\Lambda_i$, the amplitudes of the error functions, are the relative step sizes, and are chosen so that \EQ \ref{eq:comptonModelFinal} is normalized to one over the energy range 
0--20~keV. Due to the binned nature of the fit, only the amplitudes of the first four steps could be accurately determined. 
The constant term $\Lambda_0$ in  \EQ \ref{eq:comptonModelFinal} has a value of  0.005~keV$^{-1}$ and accounts for a flat background required to fit the simulated spectrum.

\begin{table}
	\centering
	\begin{tabular}{@{} p{4.4} p{4.4} p{4.4} p{4.4} @{}}
		\hline
		\multicolumn{1}{c}{$\Lambda_K$} & \multicolumn{1}{c}{$\Lambda_L$} & \multicolumn{1}{c}{$\Lambda_M$} & \multicolumn{1}{c}{$\Lambda_N$}  \\
		\hline
		5.7,0.3 & 15.2,0.5 & 9.43,1.40 & 18.7,1.3  \\
		\hline
	\end{tabular}
	\caption{Compton model parameters for CDMSlite, normalized over the energy range 0--20~keV. All values have been multiplied by a factor of 10$^3$ and are in units of keV$^{-1}$. }
	\label{tab:finalComptonFitParam}
\end{table}

\begin{figure}
	\centering
    \includegraphics[width=0.9\columnwidth]{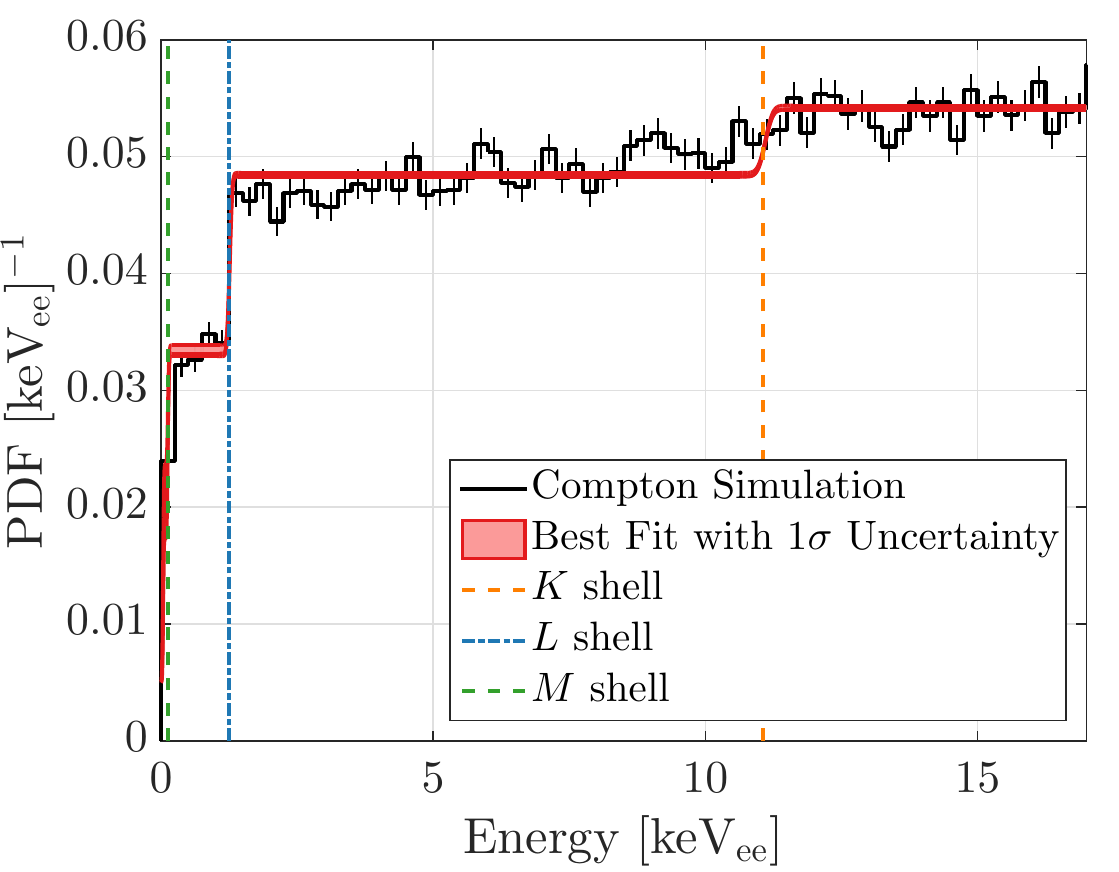}
	\caption{Best fit of the Compton scattering spectral model of Eq. \ref{eq:comptonModelFinal} to a \geant simulation of Compton scatters.}
	\label{fig:comptonT2Z1}
\end{figure}

\TAB\ref{tab:finalComptonFitParam} gives the final parameters of our Compton model, extracted from a fit of \EQ\ref{eq:comptonModelFinal} to the \geant simulation shown in \FIG \ref{fig:comptonT2Z1}.

\subsection{Surface Backgrounds}
\label{sec:surfacebackground}

Surface events are primarily due to the decay of $^{210}$Pb, which is a long-lived daughter of $^{222}$Rn. Radon exposure can cause $^{210}$Pb to become implanted into the surfaces of the detectors and their surrounding copper housings. Radiation from the $^{210}$Pb decay chain consists primarily of betas, Auger electrons, $^{206}$Pb ions, and alphas which have a small mean free path in Ge and will deposit the majority of their energy within a few millimeters of the detector's surface. To understand this background and build a model of its expected distribution in energy, we use a \geant simulation and a detector response function. We normalize the predicted rate of surface backgrounds using a study of alphas in \scdms iZIP data.

\subsubsection{Detector Response of CDMSlite}

Surface events will deposit all their energy within a few millimeters of the detector surface, depending on the particle type. Due to the asymmetric electric field shown in \FIG\ref{fig:voltageMap}, many surface events at large radii will experience reduced NTL gain and be removed by the fiducial volume cut. To properly model this background in CDMSlite, an approximation of the detector response is needed such that reduced NTL events can be removed.

The detector response model uses the voltage map of \FIG\ref{fig:voltageMap} and the resolution model of \EQ\ref{eq:simpleRes} to approximate the total phonon energy measured in the detector. Each component of the energy resolution model is implemented independently. For example, the energy deposited in a \geant simulation is used to determine the average number of electron-hole pairs produced, then an integer number of actual pairs is drawn from the distribution of width $\sigma_{\text{F}}$. A yield correction is applied to NRs based on the Lindhard model (see \EQ\ref{eq:lindhard}). The location of the \geant event in the detector is used to determine the experienced voltage $\Delta V$ for the event and thus the total phonon energy using \EQ\ref{eq:totEne_yield}. \Er is given by the energy deposited in the simulation.

We do not attempt to simulate the radial parameter $\xi$ for surface events. Instead, because the radial cut removes events at large radii that have reduced NTL amplification due to the reduced electric potential, we use a cut on the experienced $\Delta V$ of the events as a proxy for the fiducial volume cut. This was set at $\Delta V > V_{\text{cut}} \approx 74~\text{volts}$, where the simulation itself used $V_{\textrm{det}} = 75$\,volts.

\subsubsection{Simulation of $^{210}$Pb Contamination}

In \geant, we use the Screened Nuclear Recoil physics list~\cite{Mendenhall2005} to model the implantation of $^{210}$Pb into the material surfaces along with any recoil of nuclei by subsequent decays to the stable isotope $^{206}$Pb. We consider three locations from where surface events may originate: the copper directly above the detector (``top lid'', TL), the cylindrical housing (H) and the surface of the germanium crystal itself (Ge). 

We simulated energy deposition from the decays of $^{210}$Pb, $^{210}$Bi, and $^{210}$Po for the three locations. Applying the detector response function to each simulated decay yields the expected spectrum for this analysis. Additionally, we consider only events with energy deposition in the top detector of the tower (single-scatter events), since that is the location of the CDMSlite detector. The spectra from all three decays can be added under the assumption of secular equilibrium between the two daughters and the $^{210}$Pb parent. This is a valid assumption because the longest daughter half-life in this chain is 138 days, which is short compared to the time between the last exposure to radon and the beginning of the measurement. The spectra from the three locations are included in the likelihood fit of \SEC \ref{sec:likelihood:function} to account for all possible surface background events.

The voltage cut and selection of single-scatter events mimic the fiducial volume cut and multiple-scatters cut, respectively (see \SECS\ref{sec:cutOverview} and \ref{sec:fiducialvolume}). The efficiency of all analysis cuts was applied to the final simulated spectra. 

\subsubsection{Normalization}

We normalize the surface background rate with an independent measurement of the alpha decay events in the CDMSlite detector (similar to the surface-event normalization in \REF\cite{Agnese2014}), using a data set with a livetime of $\sim$380~days taken with the detector operated in iZIP mode. 
Because this iZIP-mode data set provides more detailed information on event positions, the observed rates could be attributed to surface event sources originating from parents on the top lid, housing, and detector surface. The detector surface rate is deduced from the surface facing the neighboring detector. 
This rate is then subtracted (with the appropriate surface area scaling) from the event rate measured on the side wall and the surface facing the  top lid to determine the rate from the other two locations (H and TL).
Because the determination of an individual source's contribution depends on subtracting the contribution of the other sources, this normalization procedure introduces a negative correlation between the various components.

We compare the observed alphas from the detector surface, top lid, and housing to the simulated number to determine a scaling factor for the simulation. The single-scatter events that pass the voltage cut in the simulation are then scaled to the \runThree livetime to get the expected number of surface events. The germanium, housing, and top lid are estimated to respectively contribute 3.4, 6.5, and 17 events from 0--2~\kevee after signal efficiency cuts have been applied.

\subsubsection{Discussion of Uncertainties}
\label{sec:surfacebackground:uncertainties}

There are two main sources of systematic uncertainty on the energy spectra for surface events: uncertainties in the voltage map that determines the voltage $\Delta V$ for each event, and the location of the fiducial volume cut. The map in \FIG\ref{fig:voltageMap} assumes no additional detectors in the tower. Including the detector beneath the CDMSlite detector results in a difference of 0.5\,V and 1\,V for the top and bottom faces respectively, which we incorporate as a systematic uncertainty.  Additionally, we model uncertainties in the fiducial volume cut (using the voltage cut $V_\text{cut}$ as a proxy for the radial parameter cut) by varying the voltage cut from roughly $V_{\text{cut}}-2\,$V to $V_{\text{cut}}+1\,$V. These two sources of error are independent and can be added in quadrature. 

\begin{figure*}
\subfloat{\includegraphics[width=0.662\columnwidth]{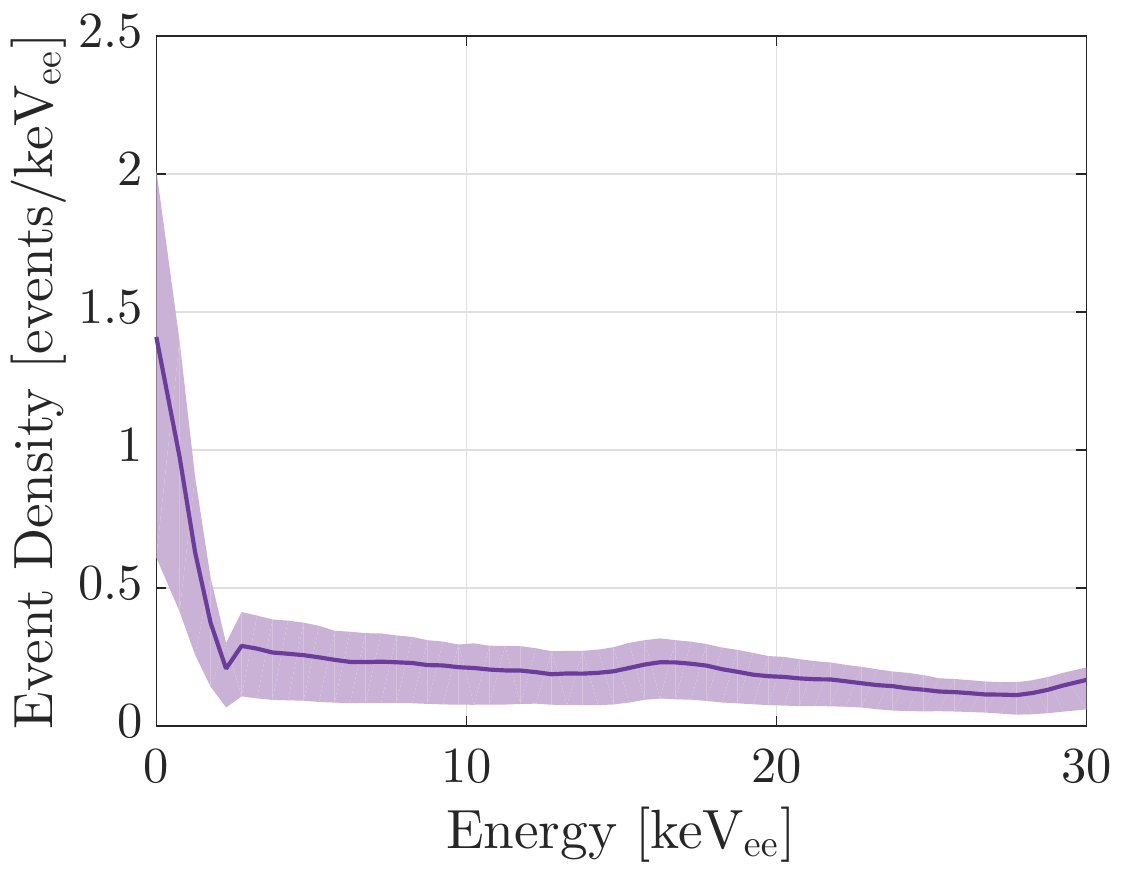}}
\hfill
\subfloat{\includegraphics[width=0.65\columnwidth]{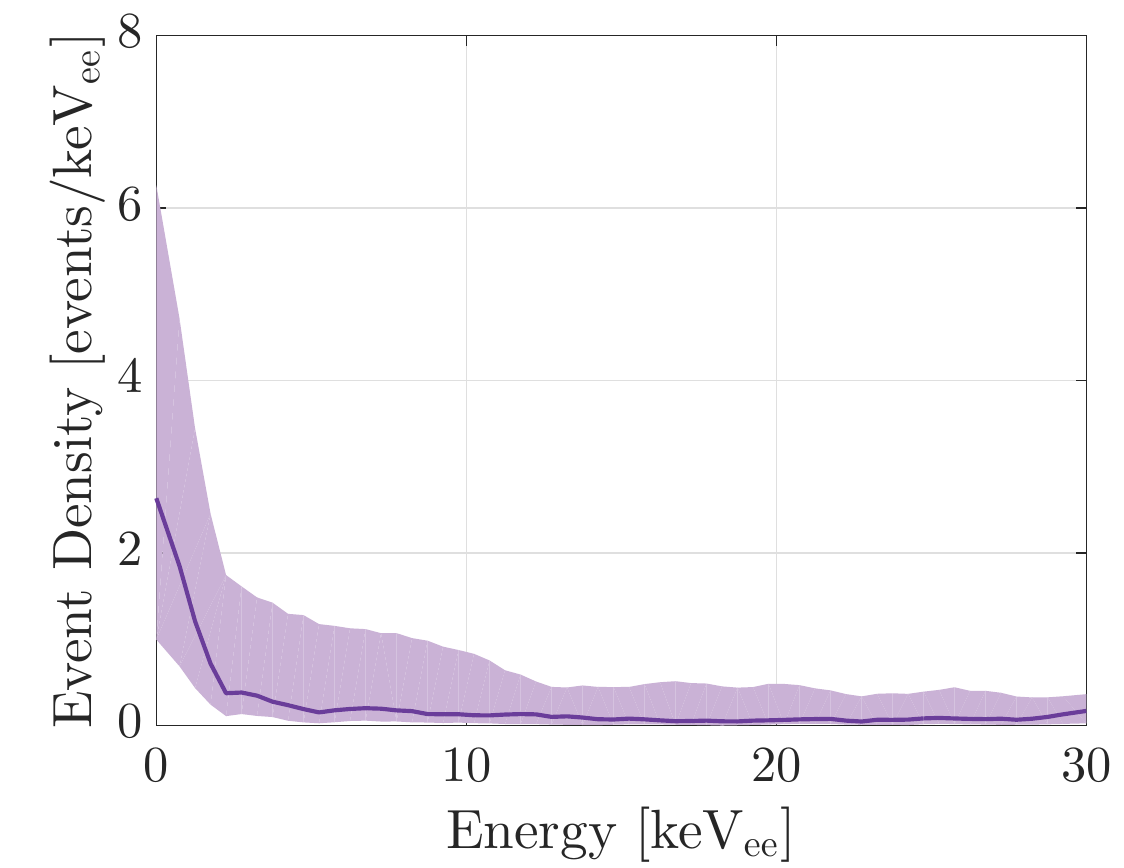}}
\hfill
\subfloat{\includegraphics[width=0.652\columnwidth]{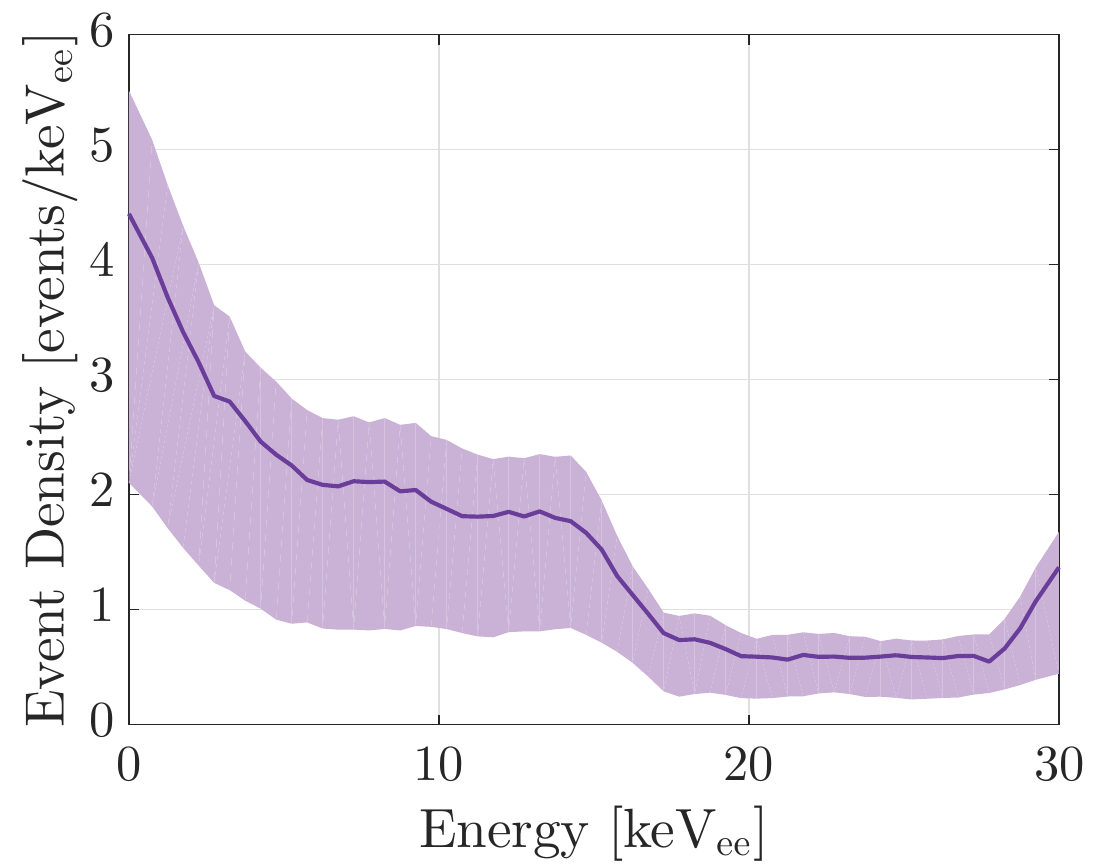}}
    \caption{The spectra (normalized to event density) of surface events expected from the three surface background locations (left: germanium; center: housing; right: top lid).  For each location, the solid curve represents the mean of the expected event distribution ($\rho_0$).  The shaded band shows the 1$\sigma$ uncertainty, where the top and bottom edges of the bands correspond to $\rho_{+}$ and $\rho_{-}$ in \EQ\ref{eq:eventdensity}, respectively.\label{fig:surfaceGeHTL}}
\end{figure*}

The surface backgrounds are included into the likelihood of \SEC\ref{sec:likelihood} using event densities, $\rho_{j}$, that are functions of morphing parameters, $m_{j}$.  Here $j$ iterates from 1 to 3, corresponding to the three surface background sources.  The morphing parameters, collectively denoted as $\vec{m}$, are used in order to incorporate both the uncertainty on spectral shape and uncertainty on the normalization from the alpha study. They allow the event density to smoothly vary within the 1$\sigma$ uncertainty band as:

\begin{equation}
\label{eq:eventdensity}
	\rho_{j}(E,m_{j}) = 
    \begin{cases}
    	\rho_{j,0} (E) + m_{j} \times \left[\rho_{j,+} (E) - \rho_{j,0} (E) \right] \\
        \rho_{j,0} (E) + m_{j} \times \left[\rho_{j,0} (E) - \rho_{j,-} (E) \right]
    \end{cases}
,
\end{equation}

\noindent
where $m_j \geq 0$ ($m_j < 0$) for the upper (lower) expression. 
A value of $m_{j} = 0$ results in the nominal event density ($\rho_{j,0}$), $m_{j} = 1$ results in the upper +1$\sigma$ event density ($\rho_{j,+}$), and $m_{j} = -1$ results in the lower $-$1$\sigma$ event density ($\rho_{j,-}$). The event densities, shown in \FIG \ref{fig:surfaceGeHTL}, are normalized such that the integral of $\rho_{j,0}$ gives the expected number of surface events as indicated by the alpha study.

Because the systematic uncertainties from the voltage cut are positively correlated between the different surface backgrounds, the morphing parameters of the three surface backgrounds are positively correlated. We encode correlations from these common systematics, as well as correlations resulting from the alpha decay normalization study, in a covariance matrix $\textbf{M}$ between morphing parameters. Information from the alpha study prefers  constraints on $\vec{m}$ centered at zero. Fits to the CDMSlite energy spectra above the region of interest for this analysis, done as part of a sensitivity study described in \SEC\ref{sec:sensitivity}, favor slightly negative values for the $\vec{m}$. We use the fitted values and covariances from that study as constraints in the likelihoood fit of \SEC\ref{sec:likelihood}.
\section{Profile Likelihood Analysis}
\label{sec:likelihood}

To incorporate information about backgrounds when searching for a DM signal, we use the profile likelihood ratio (PLR) method, which improves upon previous CDMSlite DM searches in multiple ways. First, it provides improved sensitivity over the optimum interval limit-setting method \cite{Yellin2002,Yellin2007} as implemented in the \runTwo analysis because the known backgrounds are taken into account. Second, the PLR approach can in principle be used in a discovery framework, potentially allowing for discovery of a signal. Third, the PLR approach naturally incorporates systematic uncertainties into signal and background models and reflects those systematic uncertainties in the sensitivity.

The PLR method fits the probability distribution functions (PDFs) for a DM signal and all background sources accounted for in our background model to the energy spectrum of events that pass all cuts. Separate PDFs are used for \perOne and \perTwo. CDMSlite has greatest sensitivity to DM masses between 1 and 10~\gev. Because the corresponding expected energy spectrum from a DM signal is concentrated below 2~\kevee, we restrict our final likelihood fit (and thus our DM search) to the 0.07--2~\kevee energy range, where 0.07~\kevee is the analysis threshold. Tests of the likelihood fit done prior to unsalting on simulated data sets validated the fitting method.

\subsection{Likelihood Function}
\label{sec:likelihood:function}
We use an unbinned extended likelihood to fit for the number of DM and background events in the final data set. One-dimensional PDFs, denoted by $f(E)$ and normalized to unity over the energy range of the fit, describe the signal and non-surface background distributions as a function of energy. We calculate the signal PDF using standard DM halo assumptions and the Helm nuclear form factor \cite{Lewin1996,Helm1956}, as a function of the DM mass. The number of fitted DM events is denoted $\nu_{\chi}$ and is related to the DM cross section $\sigma_{\chi}$. The non-surface background model is comprised of six PDFs from the sources discussed in \SEC\ref{sec:backgrounds}: Compton scattering events, tritium, and four different EC isotopes ($^{68}$Ge/$^{71}$Ge, $^{68}$Ga, $^{65}$Zn, $^{55}$Fe). The number of background events from these different sources is given by $\nu_{b,i}$, where $i$ iterates from 1 to 6.

We include the surface background distributions in the likelihood not as PDFs but as event densities, denoted $\rho_j(E)$, which account for both spectral shape and normalization.
This was done because the energy spectra of these backgrounds vary with the systematic uncertainties considered and correlate with their normalizations, both parameterized by the morphing parameters, $\vec{m}$, discussed in \SEC\ref{sec:surfacebackground}. The number of background events from the surface background sources is given by $\nu_{sb,j} = \int \rho_{j}(E)~dE$, where $j$ iterates from 1 to 3.

While the normalizations of the surface background event density distributions are constrained by the alpha measurements discussed in \SEC\ref{sec:surfacebackground}, we place no constraints on the number of events contributing from the other background sources. Spectral information alone is used to fit these backgrounds and differentiate them from the DM signal distribution.

The full extended likelihood function is

\begin{equation}
\label{eq:likelihood}
\begin{split}	\mathcal{L}  =  \frac{e^{-\nu_{tot}}}{N!} \prod_{i=1}^N \Big[ & \nu_{\chi}f_{\chi}\left(E_i,\vec{\alpha}\right)  + \sum_{b=1}^{6} \nu_{b,i} f_b\left(E_i,\vec{\alpha}\right) \\
& + \sum_{j=1}^{3} \rho_{j}\left(E_i,\vec{\alpha},m_{j}\right) \Big]  \times \mathcal{L}_{\text{Constr.}}(\vec{\alpha},\vec{m}),
\end{split}
\end{equation}

\noindent
where $N$ is number of events in the data set, $\nu_{tot} = \nu_{\chi} + \sum_i \nu_{b,i} + \sum_{j} \nu_{sb,j}$ is the total number of fitted signal and background events, $\vec{\alpha}$ is a set of nuisance parameters that vary the shapes of the PDFs as a function of systematic uncertainties, and $\mathcal{L}_\text{Constr.}$ is a constraint term that encodes prior constraints on these nuisance parameters as well as the morphing parameters $\vec{m}$.

\subsection{Systematic Uncertainties \& Constraints}
\label{ref:likelihood:systematics}

The $\vec{\alpha}$ parameters in \EQ \ref{eq:likelihood} incorporate systematic uncertainties from the NR ionization yield (described in \SEC\ref{sec:expDescription:energyScale}), the signal efficiency, and detector resolution into the likelihood. These sources are parametrized respectively by Lindhard's $k$ parameter, three efficiency parameters $\vec{e}$, and six  resolution parameters $\vec{r}$; so $\vec{\alpha} = \{k, \vec{e}, \vec{r}\}$. The NR ionization yield parameter $k$ shifts the signal distribution as described in \SEC\ref{sec:expDescription:energyScale}.  The signal efficiency parameters scale the distributions by the shape given by \EQ \ref{eq:efficiencyerf}, and the resolution parameters smear distributions with a resolution given by \EQ \ref{eq:simpleRes} and parameters from \TAB\ref{tab:run3resolutions}.

The $\mathcal{L}_{\text{Constr.}}$ term in \EQ \ref{eq:likelihood} is  given by

\begin{equation}
\label{eq:constraint}
\begin{split}
 \textrm{ln}( \mathcal{L_{\textrm{Constr.}}}) = & -\frac{(k - \mu_{k})^2}{2\sigma_{k}^2} \\ & - \frac{1}{2} \Big[\sum\limits_{i,j}^{3} (e_i-\mu_{e_i})\textbf{E}^{-1}_{ij}(e_j-\mu_{e_j})\Big]  
	   \\ & - \frac{1}{2} \Big[\sum\limits_{i,j}^{6} (r_i - \mu_{r_i})\textbf{R}^{-1}_{ij}(r_j - \mu_{r_j})\Big]
       \\ & - \frac{1}{2} \Big[\sum\limits_{i,j}^{3}  (m_i - \mu_{m_i})\textbf{M}^{-1}_{ij}(m_j - \mu_{m_j})\Big],
        \end{split}
	  \end{equation}
\noindent  
which constrains the $\vec{\alpha}$ and $\vec{m}$ variables by their central values, uncertainties, and correlations as determined with prior information. These constraints dictate the extent to which the systematic uncertainty parameters can alter the shape (and, in the case of $\vec{m}$, the normalization) of the signal and background distributions.

The 1D Gaussian constraint on $k$ allows this parameter's fitted value to differ from the theoretical value for germanium, $\mu_k$ = 0.157. The systematic uncertainty is the Gaussian's width, $\sigma_k$, as estimated from auxiliary measurements of the ionization yield in germanium \cite{Barker2012}. Because these measurements do not provide precise information about the NR ionization yield, particularly at low energy, we use a weak constraint on $k$ by choosing $\sigma_k = 0.05$.

We constrain the three parameters describing the signal efficiency, $\vec{e}$, with a 3D Gaussian prior using the results of \SEC\ref{sec:sigEffParam}. The center of the 3D Gaussian is given by the best-fit values of the parameters $\vec{\mu}_{e}$, and its shape is determined by the covariance matrix between best-fit values, given by $\textbf{E}$. We similarly constrain the resolution parameters, $\vec{r}$, using the 6D Gaussian prior from the resolution model of \SEC \ref{sec:energyresolutionmodel}, with best-fit resolution model values of $\vec{\mu}_{r}$ and covariance matrix $\textbf{R}$. Because the \perOne and \perTwo detector resolutions were modeled independently, $\textbf{R}$ contains zeros in elements linking the two periods. The morphing parameters, $\vec{m}$, which incorporate systematics of the surface backgrounds, are constrained in the final term of \EQ \ref{eq:constraint}. The expected values for the morphing parameters, $\vec{\mu}_{m}$, as well as the covariance matrix ($\textbf{M}$) between them, determine the constraint. We take the constraints for the morphing parameters from the sensitivity study described in Section~\ref{sec:sensitivity}.

\subsection{Upper Limit Calculation}

We test the hypothesis that a DM signal with spin-independent cross section $\sigma_{\chi}$, for a certain mass, exists in the data. Because the best-fit value of $\sigma_{\chi}$ for the DM masses considered in this analysis is found to be well below the experiment's sensitivity (calculated in \SEC \ref{sec:sensitivity}), we choose to set an upper limit. Using the likelihood ratio statistic $q_{\sigma_{\chi}}$ described in \REF\cite{XENON2011}, all parameters in the likelihood other than $\sigma_{\chi}$ (i.e. the systematic uncertainty parameters and the numbers of background events) are \textit{profiled out} as nuisance parameters by maximizing $\mathcal{L}$ as a function of these parameters with $\sigma_{\chi}$ held constant. Explicitly,  $q_{\sigma_{\chi}}$ is defined as 

 \begin{equation}
	q_{\sigma_{\chi}} = 
	\begin{cases}
		-2\text{ln} \lambda(\sigma_{\chi}) 		& \hat{\sigma}_{\chi} < \sigma_{\chi} \\
		0	& \hat{\sigma}_{\chi} > \sigma_{\chi} 
	\end{cases},
\end{equation}
 
\noindent
where $\lambda$ is defined as

\begin{equation}
	\lambda(\sigma_{\chi}) = \frac{\mathcal{L}\left(\sigma_{\chi},\hat{\hat{\theta}}\right)}{\mathcal{L}\left(\hat{\sigma}_{\chi},\hat{\theta} \right)}.
\end{equation}
The numerator of $\lambda(\sigma_{\chi})$ is the likelihood of a fit that has constrained the signal component to the test hypothesis value $\sigma_{\chi}$, and $\hat{\hat{\theta}}$ are the values of the nuisance parameters that maximize the likelihood given the constraint on $\sigma_{\chi}$. The denominator of $\lambda(\sigma_{\chi})$ is the likelihood with no constraints---the cross section $\sigma_{\chi}$ is permitted to float, along with the nuisance parameters, and the values that maximize the likelihood are labeled $\hat{\sigma}_{\chi}$ and $\hat{\theta}$. Signal hypotheses for which $\hat{\sigma}_{\chi} > \sigma_{\chi}$ are compatible with the data when calculating upper limits. Therefore $q_{\sigma_{\chi}}$ is set to 0 in these cases, which is the value that indicates the highest degree of compatibility between the signal hypothesis and the data.
 
 This profiling method yields a likelihood ratio function that is solely a function of $\sigma_{\chi}$.
 We calculate the $\sigma_{\chi}$ value for which the signal hypothesis ($H_{\sigma_{\chi}}$) is rejected at the 90\% confidence level (CL) by comparing the $q_{\sigma_{\chi}}$ obtained from the data to the expected distribution of $q_{\sigma_{\chi}}$ when the signal hypothesis is true, $g(q_{\sigma_{\chi}}|H_{\sigma_{\chi}})$. While significant computation is required to calculate $g(q_{\sigma_{\chi}}|H_{\sigma_{\chi}})$ for every tested signal hypothesis $\sigma_{\chi}$, Wilks' theorem~\cite{Wilks1938} indicates that $g(q_{\sigma_{\chi}}|H_{\sigma_{\chi}})$ asymptotically approaches a distribution that has equal contributions from a Dirac delta function distribution centered at zero and a $\chi^2$ distribution with one degree of freedom. Monte Carlo calculations have verified that $g(q_{\sigma_{\chi}}|H_{\sigma_{\chi}})$ converges to the distribution predicted by Wilks' theorem for a variety of tested signal hypotheses within the sensitivity of the \runThree analysis, and therefore the theoretical distribution is used to set the upper limit.
 
Additionally, the CL$_{s}$ technique \cite{CLS2002} is used to protect against the possibility of the PLR method excluding a DM-nucleon cross section lower than the sensitivity of the experiment, which can occur if the background statistically fluctuates to a low number of events. A consequence of this protection, which we have verified with Monte Carlo simulations, is that the CL$_s$ technique gives a slightly higher 90\% excluded signal cross section than would otherwise be obtained (i.e. provides a limit with over-coverage) and is therefore conservative.

\section{Results}
\label{sec:results}

\subsection{Sensitivity Calculation}
\label{sec:sensitivity}

Prior to unsalting the data, we calculated the 90\% CL sensitivity of the \runThree analysis to a DM signal based on projected background rates in this analysis's energy region of interest (ROI), 0.07--2.0~\kevee. The sensitivity calculation also uses the likelihood framework presented in \SEC \ref{sec:likelihood}. To estimate the background rates in the ROI, we measure them in the 5--25~\kevee range and extrapolate the rates to lower energy. We choose 5~\kevee because salt was not inserted above this energy and because the DM signal contribution above this energy for DM masses $<$\,10~\gev is expected to be negligible. Also, because 5~\kevee is below the lowest $K$-shell energy of the EC isotopes considered, all background components are constrained in this range.
We perform a maximum likelihood fit, using the likelihood defined in \EQ \ref{eq:likelihood} but without the DM signal.  We also omit the resolution and efficiency systematic uncertainties because those extra terms are unnecessary when fitting the 5--25~\kevee background spectrum.  This fit provides best-fit values of, as well as covariances between, background rates in the 5--25~\kevee range for the nine background components. The expected background in the ROI can directly be calculated from the best fit in the 5--25~\kevee range. The uncertainty is determined from the covariance matrix of the fit.

Background-only pseudo-experiments are then generated by sampling from the nine different background distributions. The number of events thrown for each background component is randomized, first by sampling from the 9D Gaussian distribution provided by the 5--25~\kevee maximum likelihood fit and second by adding a Poissonian fluctuation to the sampled value. The 90\% CL PLR limit, using the CL$_s$ technique, is calculated for 500 of these pseudo-experiments, and the resulting $\pm$ 1$\sigma$ and $\pm$ 2$\sigma$ sensitivity bands are shown respectively by the green and yellow bands in \FIG \ref{fig:lim1}.

In addition to determining parameters for generating the pseudo-experiments, the 5--25~\kevee fit provides constraints on the surface background morphing parameters (the $\mu_{m_i}$ of \EQ\ref{eq:constraint}). While this fit used a prior constraint centered at 0 for all morphing parameters, the respective posteriors peaked at $-$0.19, $-$0.2, and $-$0.25 for the germanium, top lid, and housing surface background locations respectively. This indicates a slightly lower surface background rate than predicted by the alpha decay study.  These updated central values for the constraint were used in the likelihood for both the sensitivity estimate and the final limit, along with an updated covariance matrix for the morphing parameters.

\begin{figure}
	\centering
    \includegraphics[width=\columnwidth]{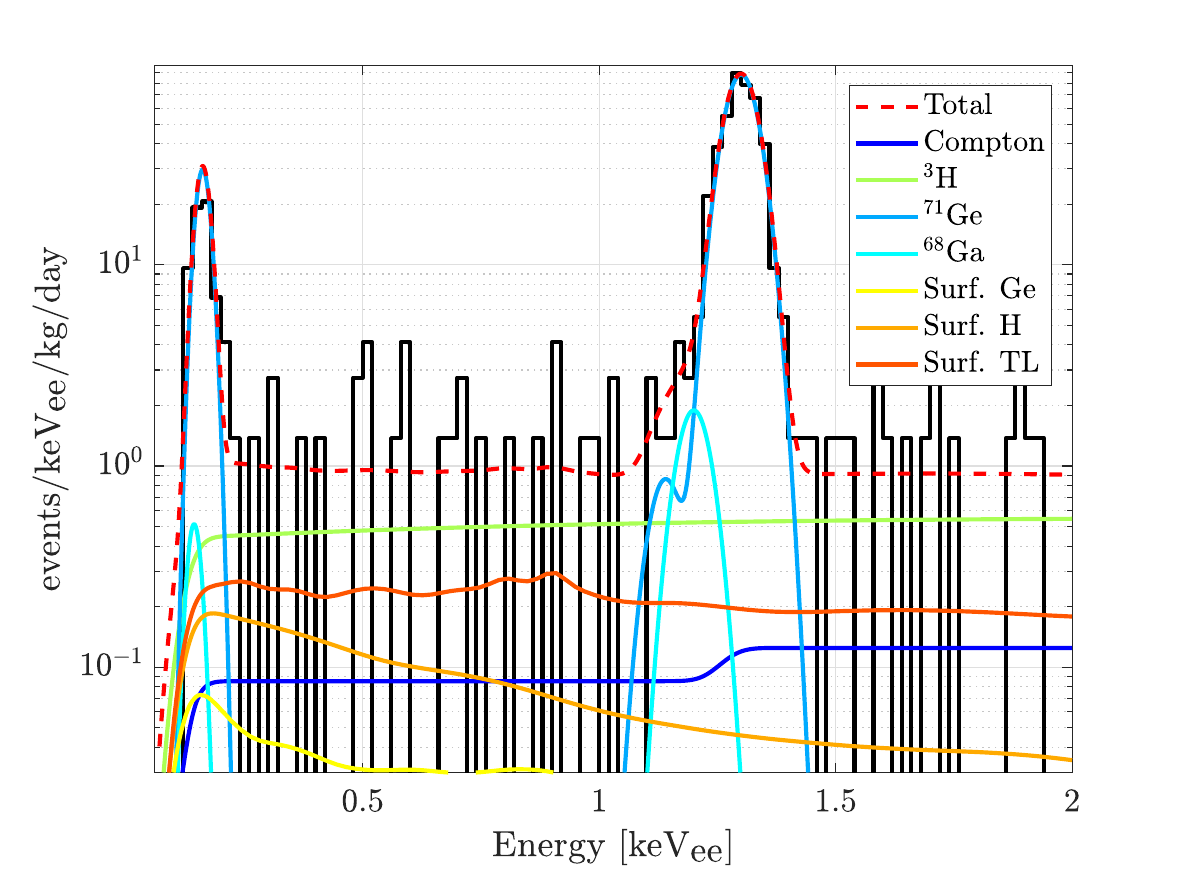}
	\caption{The CDMSlite \runThree final energy spectrum overlaid with the best-fit background components. The best-fit rates for the $^{65}$Zn and $^{55}$Fe components are below the scale of the plot.} 
	\label{fig:fittedSpec}
\end{figure}

\subsection{Evaluating the Goodness of Fit}

Because the likelihood fitting procedure described in \SEC \ref{sec:likelihood:function} provides no information as to the goodness of fit (GOF) of the model to the data, we define a procedure to evaluate the GOF that estimates a probability (i.e. a $p$-value) for the data given the model. We use the Cram\'er-von Mises GOF statistic \cite{Birnbaum1953} because it does not require binning of the data, overcomes some deficiencies of the more common Kolomogorov-Smirnov test, yet is still relatively simple compared to some alternative GOF metrics.

The particular GOF procedure that we use incorporates the systematic uncertainties described in \SEC \ref{ref:likelihood:systematics}. We fit the data using the likelihood of \EQ \ref{eq:likelihood} without a DM component and calculate the Cram\'er-von Mises statistic using the best-fit total background model (i.e. the red dashed line in \FIG \ref{fig:fittedSpec}). One output from the fit to the data is the covariance between all systematic uncertainty parameters. We then generate 1000 pseudo-experiments that are representative of the model's fit to the data. Using the covariance matrix between systematic uncertainty parameters, we randomize the systematic uncertainty parameters for each pseudo-experiment, which slightly changes the shape of the individual background components. We then sample those individual background components using Poisson fluctuations around the best-fit value from the fit to the final spectrum. Finally, we fit these pseudo-experiments and calculate a Cram\'er-von Mises statistic for each of them. The $p$-value is then the fraction of pseudo-experiments with a Cram\'er-von Mises statistic greater than the one for the data fit.

Prior to unsalting, we agreed on a $p$-value threshold of 0.05, below which we would investigate inaccuracies in the background model, abandon the limit obtained with the profile likelihood method, and resort to the more conservative optimum interval \cite{Yellin2002,Yellin2007} limit-setting technique. Upon unsalting we found a $p$-value of 0.988, indicating a particularly good fit. Checks of biases in the GOF evaluation were performed and none were discovered. We therefore accept the 90\% CL limit provided by the profile likelihood method.

\subsection{DM Limit and Background Rates}

The final \runThree spectrum after application of all selection criteria is shown in \FIG \ref{fig:fittedSpec}. The main features are the $^{71}$Ge electron-capture $L$- and $M$-shell peaks at 1.30 and 0.16~\kevee respectively. Events contributed from backgrounds other than $^{71}$Ge exist between the peaks and are well modeled. We do not observe a population of events below the $M$ shell, which is consistent with the steep decrease of the signal efficiency in this range and consistent with the expectations from the background model. 

While the best-fit individual background components are shown in \FIG \ref{fig:fittedSpec}, this figure does not provide a visualization of the covariances between background components. As expected, a strong covariance is observed between the Compton and $^3$H background components, which in this energy range do not contain sufficiently distinct spectral features to remove their degeneracy in the fit. The surface background components are strongly correlated through the prior constraint covariance matrix, $\textbf{M}$, described in \SEC \ref{sec:surfacebackground:uncertainties}. We find that the surface background component covariances from the likelihood fit match the prior constraint covariances, indicating that these 0.07--2.0~\kevee data do not provide any additional information on the surface background.

\begin{table}
	\centering
	\begin{tabular}{h{4.3}p{3.3}p{3.3}}
		\hline \hline
		\multicolumn{1}{c}{Range}	& \multicolumn{1}{c}{\runTwo Rate}	& \multicolumn{1}{c}{\runThree Rate }	\bigstrut[t] \\
		\multicolumn{1}{c}{$\left[\keveeeq\right]$}	& \multicolumn{1}{c}{$\left[\keveeeq\,\text{kg}\,\text{d}\right]^{-1}$}	& \multicolumn{1}{c}{$\left[\keveeeq\,\text{kg}\,\text{d}\right]^{-1}$} \bigstrut[b] \\
		\hline
		0.2,1.2	& 1.09,0.18			& 1.9,0.3			\\
		1.4,10	& 	1.00,0.06		& 1.3,0.1			\\
		11,20		&	0.30,0.03		& 0.71,0.07	\bigstrut[b]	\\
		\hline \hline
	\end{tabular}
	\caption{Average single-scatter event rates for energy regions between the activation lines in \runTwo and \runThree, corrected for efficiency. All errors contain ${\pm}\sqrt{N}$ Poissonian uncertainties, and the lowest energy range values additionally include uncertainty from the signal efficiency.}
	\label{tab:rawRates}
\end{table}

We calculate the average background rates of single-scatter events between the $^{71}$Ge peaks, corrected for efficiency, as shown in \TAB \ref{tab:rawRates}. 
 The higher background rates, relative to \runTwo, are consistent with the expected background rates based on the position of the detector in the tower. The \runTwo detector had neighboring detectors on both of its faces. By contrast, the \runThree detector was the top detector in the tower and therefore had one face exposed to the top copper lid. Additionally, it is expected that identification of multiple scatters in the \runThree detector is diminished because of its position in the tower; therefore, a higher fraction of multiple scatter events could be passing the multiples cut and contributing to the background rates shown in \TAB \ref{tab:rawRates} for \runThree.

\begin{figure}
	\centering
    \includegraphics[width=\columnwidth]{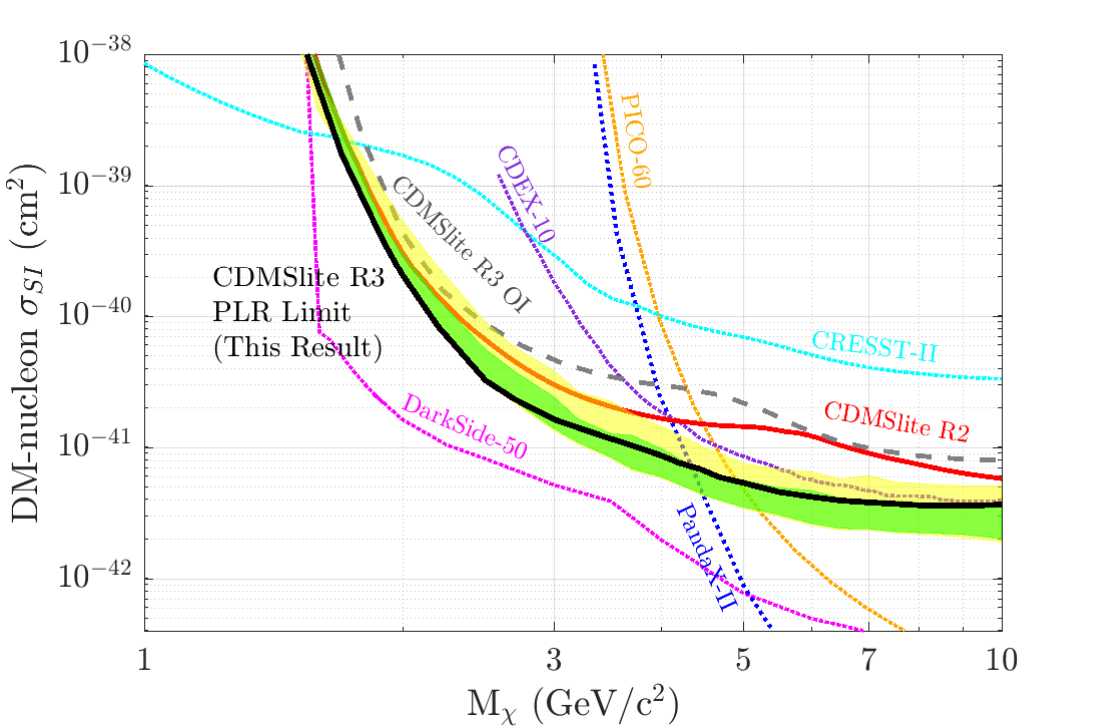}
	\caption{The CDMSlite \runThree 90\% CL PLR limit (this result, solid black) on the spin-independent WIMP-nucleon cross section, along with the $\pm$\,1$\sigma$ and $\pm$\,2$\sigma$ sensitivity bands (green and yellow respectively). The CDMSlite \runThree optimum interval limit (dashed grey) and \runTwo (red) optimum interval limit \cite{Agnese2016} are overlaid. Examples of limits from other detector technologies are overlaid: DarkSide-50 2018 No Quenching Fluctuations (magenta)~\cite{Agnes2018}; PandaX-II 2016 (blue)~\cite{pandax2016}; PICO-60 2017 (orange)~\cite{pico2017}; CRESST-II 2016 (cyan)~\cite{cresst2016}; CDEX-10 2018 (purple)~\cite{cdex2018}.}
	\label{fig:lim1}
\end{figure}

Figure \ref{fig:lim1} shows the final CDMSlite \runThree limit calculated with the spectrum in \FIG \ref{fig:fittedSpec}. From 2.5--10 \gev we find a factor of 2--3 improvement in the excluded DM-nucleon cross section over the CDMSlite \runTwo optimum interval analysis \cite{Agnese2016}. This improvement is achieved despite the smaller exposure (36 vs.\ 70 kg-days) and higher background rate in \runThree, demonstrating the discrimination power of the PLR method. Below 2.5~\gev, we exclude little to no additional parameter space because the effective energy threshold for this analysis is slightly higher than that for CDMSlite \runTwo.

\section{Summary}
\label{sec:summary}

These results demonstrate successful modeling of radioactive backgrounds in CDMSlite detectors down to very low energies, as well as the power of a profile likelihood fit to set strong limits on a potential DM signal even in the presence of irreducible backgrounds. This analysis sets an upper limit on the dark matter-nucleon scattering cross section in germanium of 5.4$\times$10$^{-42}$\,cm$^2$ at 5\,GeV/$c^2$\xspace, which is a factor of $\sim$2.5 improvement over the previous CDMSlite result. Unlike previous CDMSlite analyses, the profile likelihood method used here potentially permits the detection of a signal. Key analysis developments enabling this approach include improved rejection of instrumental backgrounds using detector-detector correlations in a boosted decision tree, removal of events at high radii with misreconstructed energies by an improved fiducial volume cut, and Monte Carlo modeling of surface backgrounds in the detector.  The SuperCDMS collaboration is currently constructing a new experiment, SuperCDMS SNOLAB, which will use the CDMSlite
technique in detectors designed specifically
for high-voltage operation~\cite{Kurinsky2016,Agnese2017}. The results obtained here provide a proof of principle that backgrounds for these detectors can be successfully understood at a level that would permit not merely the setting on upper limits in the presence of backgrounds, but potentially the discovery of a low-mass DM signal.

The SuperCDMS collaboration gratefully acknowledges technical assistance from the staff of the Soudan Underground Laboratory and the Minnesota Department of Natural Resources. The iZIP detectors were fabricated in the Stanford Nanofabrication Facility, which is a member of the National Nanofabrication Infrastructure Network, sponsored and supported by the NSF. Funding and support were received from the National Science Foundation, the U.S. Department of Energy, Fermilab URA Visiting Scholar Grant No.\ 15-S-33, NSERC Canada, the Canada Excellence Research Chair Fund, and MultiDark (Spanish MINECO). The SuperCDMS collaboration prepared this document using the resources of the Fermi National Accelerator Laboratory (Fermilab), a U.S. Department of Energy, Office of Science, HEP User Facility. Fermilab is managed by Fermi Research Alliance, LLC (FRA), acting under Contract No. DE-AC02-07CH11359. Pacific Northwest National Laboratory is operated by Battelle Memorial Institute under Contract No. DE-AC05-76RL01830 for the U.S. Department of Energy. SLAC is operated under Contract No. DEAC02-76SF00515 with the U.S. Department of Energy.



\bibliography{refs}
\bibliographystyle{apsrev4-1-JHEPfix-autoEtAl}

\end{document}